\documentclass{pasj00}
\draft

\usepackage{natbib}
\usepackage{color}
\newcommand{\feoh}{[{\rm Fe} / {\rm H}]}
\newcommand{\abra}[2]{{\rm {#1}}/{\rm {#2}}}
\newcommand{\msun}{\, M_\odot}
\newcommand{\nuc}[2]{${}^{#1}{\rm {#2}}$}
\newcommand{\nucm}[2]{{}^{#1}{\rm {#2}}}
\newcommand{\dmix}{$\Delta ({}^{13}{\rm C}/{}^{12}{\rm C})$}
\newcommand{\dmixm}{\Delta ({}^{13}{\rm C}/{}^{12}{\rm C})}
\newcommand{\heconv}{\rm He\mathchar`-conv}

\begin{document}
\SetRunningHead{Nishimura et al.}{Light Element Synthesis in Extremely Metal-Poor AGB Stars}
\Received{2008/06/05}%{yyyy/mm/dd}
\Accepted{2009/05/21}%{yyyy/mm/dd}

\title{Oxygen And Light Element Synthesis by Neutron-Capture Reactions in Metal-Free And Extremely Metal-Poor AGB Stars}

\author{Takanori \textsc{Nishimura} 
  \thanks{JSPS Research Fellow, Present Address : Astronomical Data Archives Center, National Astronomical Observatory of Japan, Mitaka, Tokyo 181-8588, Japan}}
\affil{Department of Physics, Faculty of Science, Hokkaido University, Sapporo, Hokkaido 060-0810, Japan}
\email{nishimura.takanori@nao.ac.jp}

\author{Masayuki \textsc{Aikawa}
  \thanks{Institut d'Astronomie et d'Astrophysique, C.P.226, Universit\'e Libre de Bruxelles, B-1050 Brussels, Belgium}}
\affil{Hokkaido University OpenCourseWare, Hokkaido University, Sapporo, 060-0811, Japan}
\email{aikawa@ec.hokudai.ac.jp}

\author{Takuma {\sc Suda}
  \thanks{Astrophysics Group, EPSAM, Keele University, Keele, Staffordshire ST5 5BG, UK}}
\affil{Department of Physics, Faculty of Science, Hokkaido University, Sapporo, Hokkaido 060-0810, Japan}
\email{suda@astro1.sci.hokudai.ac.jp}

\author{Masayuki Y. {\sc Fujimoto}}
\affil{Department of Physics, Faculty of Science, Hokkaido University, Sapporo, Hokkaido 060-0810, Japan}
\email{fujimoto@astro1.sci.hokudai.ac.jp}


\maketitle

\begin{abstract}
	The metal-free (Pop.~III) and extremely metal-poor (EMP) stars of low- and intermediate-masses experience mixing of hydrogen into the helium convection during the early TP-AGB phase, differently from the meal-rich stars. 
We study the nucleosynthesis in the helium convective zone with \nuc{13}C formed from mixed protons as neutron source by using a nuclear network from H through S. 
   In the absence or scarcity of the pristine metals, the neutron-recycling reactions, $\nucm{12}{C} (n, \gamma) \nucm{13}{C} (\alpha, n) \nucm{16}{O}$ and also $\nucm{16}{O} (n, \gamma) \nucm{17}{O} (\alpha, n) \nucm{20}{Ne}$ promote the synthesis of O and light elements, including their neutron-rich isotopes and the odd atomic number elements.   
   Based on the results, we demonstrate that the peculiar abundance patterns of C through Al observed for the three most iron-deficient, carbon-rich stars can be reproduced in terms of the nucleosynthesis in Pop.~III, AGB stars in the different mass range.  
   We argue that these three stars were born as the low-mass members of Pop.~III binaries and later subject to the surface pollution by the mass transfer in the binary systems.  
   It is  also shown that the AGB nucleosynthesis with hydrogen mixing explains the abundances of C, O, Na, Mg and Al observed for most of carbon-enhanced EMP (CEMP) stars, including all CEMP-$s$ stars with s-process elements. 
   In addition the present results are used to single out other nucleosynthetic signatures of early generations of stars.
\end{abstract}

\section{Introduction}

During the past decade, the studies of extremely metal-poor (EMP) stars in the Galactic halo with high resolution spectroscopy have revealed peculiar abundance patterns, different from those known from previous observations and predicted from theoretical calculations.  
   Striking examples are the two stars of giant HE0107-5240 \citep{Christlieb2002} and sub-giant HE1327-2326 \citep{Frebel2005,Aoki2006}, which have the smallest iron abundances among the known EMP stars, i.e., of $\feoh= -5.3 $ and $-5.4$, respectively.  
   The both stars share the characteristic feature among the light element abundances; 
   apart from the huge enrichment of carbon ($[\abra{C}{Fe}] = 4.0$ and 4.26), the light elements such as nitrogen, oxygen, sodium, magnesium, and aluminum are reported to be largely enhanced \citep{Christlieb2002,Christlieb2004,Bessell2004,Frebel2005,Frebel2006,Aoki2006}. 
   Some differences are also discerned between these two stars;  
   the enhancements of nitrogen, oxygen, and Na through Al are much larger for HE1327-2326 than for HE0107-5240 (by 2.2 dex, 0.5 dex, and $1.2 - 1.5$ dex, respectively). 
   Moreover, the former displays the enrichment of an s-process element, strontium ($[\abra{Sr}{Fe}] = 1.1 \sim 1.3$), while the latter does not ($[\abra{Sr}{Fe}] < -0.5$).  
   Recently, another giant HE0557-4840 is identified at the iron abundance of $\feoh = -4.75$, much smaller iron abundance than other EMP stars of $\feoh \gtrsim -4$ \citep{Norris2007}, which also has a large carbon enrichment but smaller than the above two stars ($[\abra{C}{Fe} ] = 1.65 - 2.08$).  

In the theoretical point of view, new or improved models of stellar evolution and nucleosynthesis are required to explain such extraordinary abundance patterns observed, including both the similarities and differences.
   In this paper, we call the two stars of $\feoh < -5$ and the other star of $\feoh <-4$ as hyper metal-poor (HMP) and ultra metal-poor (UMP) stars, respectively, as proposed by \citet{Beers2005} to distinguish them from the other EMP stars of $\feoh \gtrsim -4$. 
   Several theoretical approaches have been made to understand the abundance patterns observed for these two HMP stars.  
   A straightforward interpretation of the observed small iron abundances is that these stars are the second-generation stars, born from gas polluted by the ejecta of first generation supernova. 
   \citet{Umeda2003} and \citet{Iwamoto2005} proposed a peculiar supernova model of metal-free $25 \msun$ star that gives an ejecta of unusual abundances with little iron and large amount of CNO elements, similar to those observed for HE-0107-5240 and HE1327-2326, respectively \citep[see also]{Tominaga2009}.  
   \citet{Limongi2003} presented a scenario involving two metal-free supernovae, one with low mass ($\sim 15 \msun$) and the other with high masses ($\sim 35 \msun$) for HE0107-5240. 
   \citet{Meynet2006} argued a combination of stellar wind and supernova ejecta of rotating massive star for HE1327-2326.  
   For all these scenarios adopt the crude supernova ejecta, however, the oxygen abundance may be a critical observation in discriminating possible scenarios as pointed out by \citet{Bonifacio2003}. 
   Actually, \citet{Bessell2004} and \citet{Frebel2006} derive rather small oxygen enrichment in disagreement with these supernova scenarios. 
 
   The EMP stars, currently observed in the Galactic halo, are the low-mass stars and have survived their long lives to date.
   We should take into account the modifications of their surface abundances through not only the internal but also the external processes after their birth.
	As discussed in the previous paper \citep{Suda2004}, the origin of iron-group elements with such tiny abundances can readily be explained in terms of the accretion of the interstellar gas, enriched with metals ejected from the supernovae of later generations, in the clouds where they were born.  
   In addition, it is also argued that the enhancements of carbon and light elements are consequent upon the surface pollution by mass transfer from an AGB companion in the binary systems of metal-free stars.  
   The latter possibility attracts our interest because it suggests that these stars may belong to the first generation stars, which are called as Population III (Pop III).  
   The stars that have formed just after the Big Bang can provide information on the formation of first-collapsed objects and on the chemical and dynamical evolution in the very early universe.  

In order to discuss the possibility that the HMP and UMP stars are low-mass survivors of Pop III stars, we have to develop new aspect of evolution of extremely metal-poor stars.  
For the metal-free and extremely metal-poor (EMP; $\feoh \lesssim -2.5$) stars, it is proved that the helium-flash convection extends into the hydrogen-rich envelope in the stars of mass $M \lesssim 3.0 \msun$, different from metal-rich populations \citep{Fujimoto1990,Fujimoto2000}.  
   What causes this difference is a lower entropy of the hydrogen-burning shell in the stars of smaller CNO abundances, and hence, a smaller entropy barrier between the helium-zone and the tail of hydrogen-rich envelope. 
   The engulfed hydrogen is carried down by convection and captured by \nuc{12}C in the middle of helium convective zone to create \nuc{13}C through the $\beta$-decay of \nuc{13}N.  
   Then \nuc{13}C is carried further inward to emit the neutrons through the $\nucm{13}{C} (\alpha, n) \nucm{16}{O}$ reaction, which initiates the neutron capture reactions and the s-process nucleosynthesis. 
   The nuclear products in the helium convective zone will later be brought up to the surface, dredged up by the envelope convection that deepens in mass as the burning shells expand absorbing the energy released during the shell flashes \citep{Hollowell1990, Fujimoto2000,Suda2004,Iwamoto2004}.  

The nucleosynthesis triggered by the \nuc{13}C-burning in the helium-flash convection was originally proposed by \citet{Sanders1967}, relying upon an early computation of helium shell flashes by \citet{Schwarzschild1967}, which met with hydrogen mixing into the helium convection for $1 \msun$ stars of Population II composition.  
   \citet{Ulrich1973} elaborated this line of thought to propose a plum mixing model and argued the repeated irradiation of neutrons during the recurrence of helium shell flashes \citep[see also][]{Scalo1973}.  
   But it has been later demonstrated that the entropy barrier becomes too high and prevents the helium flash convection from reaching the tail of hydrogen-rich envelope if the radiation pressure is properly taken into account in the Pop.\ I and II stars \citep{Iben1976, Fujimoto1977}.  
   \citet{Iben1982} then argued that hydrogen could be carried down to form \nuc{13}C by the semi-convection which developed in the carbon-rich helium zone during the third dredge-up ;   
   they propose that the latter acts as neutron source as it is incorporated into the helium convection during the next shell flash, but later, \citet{Straniero1995} found that \nuc{13}C burned during the inter-pulse phase.  
   On the other hand, it is shown that for very thin envelope, the entropy of envelope decreases to allow the helium flash convection to extend through hydrogen-rich layer \citep{Fujimoto1977}.  
   Finally, it is revealed that in the extremely metal-poor stars, the entropy of hydrogen-rich envelope can be small enough to mix hydrogen into the helium flash convection \citep{Fujimoto1990, Fujimoto1995, Fujimoto2000}.    

   The purpose of the present work is to explore the characteristics of nucleosynthesis in the helium-flash convective zone, triggered by the hydrogen injection in the extremely metal-poor stars of low and intermediate masses. 
   It has been suggested that the neutron-recycling reactions, $\nucm{12}{C} (n, \gamma) \nucm{13}{C} (\alpha, n) \nucm{16}{O}$, play a key role in producing oxygen \citep{Gallino1988, Jorissen1989} since neutrons, emitted in the helium-flash convection, may be absorbed by \nuc{12}C because of the large abundance.  
   A definite conclusion could not be reached in those days, however, because of large uncertainties of neutron capture cross sections of \nuc{12}C and \nuc{16}O.  
   Recently, the neutron capture rates for these nuclei have been improved \citep[][and also see Bao et al. 2000]{Ohsaki1994,Igashira1995}, and this enables us to draw a more quantitative and sound picture.  
   Under the initial scarcity of neutron poisons in the extremely metal-deficient stars, the neutron recycling reactions are expected to have critical importance, and may further lead to the generation of heavier elements through the sequences of neutron and $\alpha$-captures.  
   In the present paper, we focus on the progress of neutron recycling reactions and the resultant formation of light elements from oxygen through sulfur including their isotopes as well as odd number nuclei of sodium, aluminum and phosphorus.  
   The synthesis of heavy elements beyond sulfur and of the s-process elements will be discussed in a subsequent paper (Nishimura et al., in preparation).  
   Since the nuclear products, thus synthesized in the helium convection, are eventually carried out to the surface, the present results may have relevance to the observed abundances for the extremely metal-poor stars.
   In particular, the comparison with the abundance patterns observed from the two most metal-deficient stars, HE 0107-5240 and HE 1327-2326 may give an insight into their origins.  

The convective \nuc{13}C-burning model, studied in this paper, is different from the so-called radiative \nuc{13}C-burning model. 
   The former is founded on the evolution of EMP stars within the current standard framework of stellar evolution. 
   In contrast, the latter resorts to the alleged formation of \nuc{13}C-pocket in the radiative helium zone during the third dredge-up;
   hydrogen is assumed to be injected from the bottom of surface convection into the top of helium zone where a thin \nuc{13}C-rich layer is formed via the proton-capture of \nuc{12}C.  
   The formation of \nuc{13}C-pocket is presumably due to the convective overshooting, though it has not yet established.
   If it is properly formed, \nuc{13}C burns to emit neutrons and trigger the s-process nucleosynthesis in the radiative zone at relatively low temperatures before the next helium flash ignites \citep[see the review by][]{Busso1999}.  
   It has been attempted to apply the radiative \nuc{13}C-burning model to the s-process nucleosynthesis in a zero-metal AGB stars \citep{Goriely2001} and in very metal-poor AGB stars \citep[$Z = 5 \times 10^{-5}$,][]{Straniero2004}.  

We describe the methods and assumptions of the present work in the next section.  
   Results of the nucleosynthesis in the helium flash convective zone, computed with the adopted nuclear reaction network, are presented and examined in Section~3. 
   In Section~4, the characteristics of their yields are examined and compared with the observations for HMP/ UMP stars and EMP stars on the production of carbon and oxygen and also on the relative abundances of sodium, magnesium, and aluminum.  
   Then, we inquire into the nature of currently known, three EMP/UMP stars. 
   The conclusions and discussion follow. 

\section{Methods and Approximations}

\begin{figure}
\FigureFile(120mm,120mm){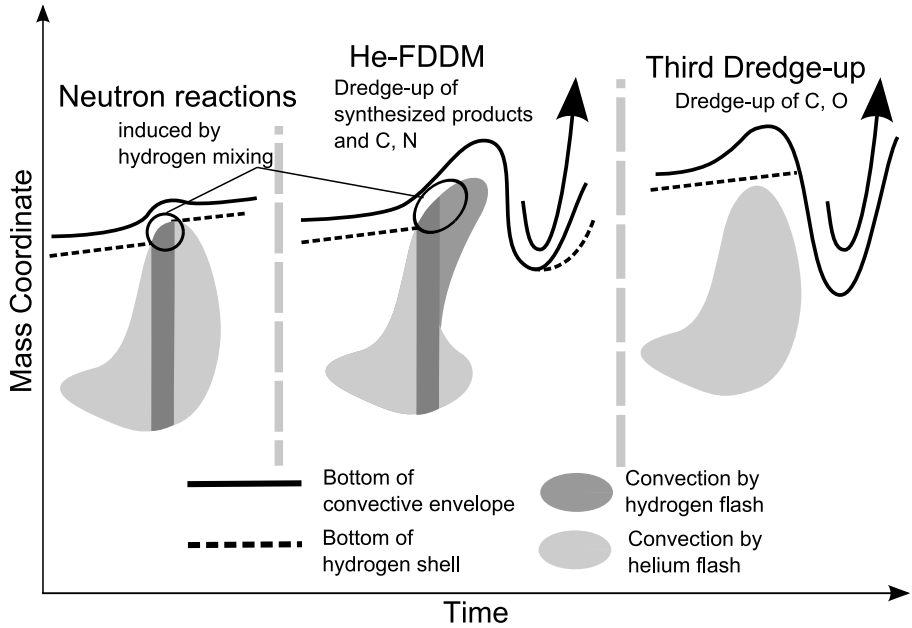}
\caption{
A schematic drawing of the progress of mixing during helium shell flashes in extremely metal-poor stars. 
   The convection driven by a helium flash extends outwards beyond the bottom of hydrogen-containing layer; 
   mixed hydrogen is carried down into the convective zone to be captured by \nuc{12}C and to form \nuc{13}C after $\beta$-decay. 
   \nuc{13}C is further mixed down to the bottom of helium convective zone and release neutrons to initiate neutron-capture reactions and synthesize s-process nucleosynthesis (left panel). 
   When the helium shell flash grows strong and mixes hydrogen at a sufficiently large rate, eventually, the burning of mixed hydrogen splits the flash convective zone into two, i.e., an upper one driven by the hydrogen shell flash and a lower one driven by the helium shell flash.  
  During the decay phase of the shell flash, the surface convection penetrates into the shells involved in the upper convective zone to cause helium-flash driven deep mixing (He-FDDM) and carry out nuclear products to the surface (middle panel).  
  For stars of $M \gtrsim 1.5 \msun$, third dredge-up events (TDU) occur during the following shell flashes.
  Carbon and oxygen, newly synthesized by 3$\alpha$ reaction and $\alpha$-capture of \nuc{12}C during the helium shell flashes without the hydrogen mixing, are dredged up to the surface along with the remainder of nuclear products, formed by the former neutron capture burning and stored in the helium zone (right panel).  
   For the detailed description about the model, see the following references \citep{Hollowell1990, Fujimoto2000, Iwamoto2004, Suda2004, Komiya2007}.
}
\label{mixing}
\end{figure}

A distinctive feature in the evolution of extremely metal-poor stars ($\feoh \lesssim -2.5 $) is that the helium convection penetrates into the hydrogen-rich layer and mixes hydrogen down into the convective zone \citep{Fujimoto2000}.  
   This occurs for the stars of mass $M \lesssim 3 \msun$. 
   The progress of nucleosynthesis and material mixing during the helium shell flashes is schematically illustrated in Figure~\ref{mixing}.   
   As the TP-AGB phase starts, the helium shell flashes recur and grow in strength to drive convection more outward, until it reaches the hydrogen-rich layer \citep{Fujimoto1990,Fujimoto2000}.  
   Hydrogen mixed into the helium-flash convection is captured by \nuc{12}C to form \nuc{13}{C} via the reactions $\nucm{12}{C} (p, \gamma) \nucm{13}{N} (e^+ \nu) \nucm{13}{C}$ in the middle of convective zone. 
   It diffuses further into deeper interior and releases neutron via $\nucm{13}{C} (\alpha, n) \nucm{16}{O}$ reaction to promote the nucleosynthesis via neutron capture reactions.   
   These nuclear products are stored in the helium zone during the recurrence of shell flashes.   
   Finally, when the helium flash grows strong to erode hydrogen at a sufficiently large rate, the burning of mixed hydrogen ignites a hydrogen-shell flash and splits the helium convective zone into two in the middle.
   This brings about the helium flash-driven deep mixing (He-FDDM) that conveys the nuclear products, stored in the helium zone and processed by hydrogen burning, to the surface \citep{Hollowell1990,Iwamoto2004}.   
   After He-FDDM enriches the envelope with CN elements above $[\abra{CNO}{H}] \simeq -2.5$, the hydrogen mixing can no longer take place because of a large entropy barrier between the helium flash-convective zone and the bottom of hydrogen-rich envelope.  
   If the star is more massive than say, $M \sim 1.5 \msun$, the third dredge-up follows to bring up to the surface the nucleosynthetic yields in the helium flash-convective zone, including freshly synthesized \nuc{12}C.   
   The s-process elements are also dredged up to the surface, as will be discussed in the subsequent paper.  
   The detailed description about the model is given in the previous papers \citep{Fujimoto2000, Suda2004, Komiya2007}.  

We investigate the progress of nucleosynthesis in the helium-flash convection considering neutron production through the hydrogen mixing.  
   The variations of the strength of shell flashes are provided by the simulations of hydrogen mixing performed with stellar evolution codes for various initial masses and pristine metallicity \citep[see][ and the references therein and Campbell \& Latanzio 2008]{Suda2008a}.
   Furthermore, the efficiency of elemental mixing is subject to uncertainties due to the treatment of convection and the entrainment of matter by convective elements.  
   Since the purpose of this work is to explore the general characteristics of nucleosynthesis triggered by hydrogen mixing, we may well treat the amount of mixed hydrogen and the duration of mixing event as parameters.   
   In our computations, we deal with the mixing of \nuc{13}C, instead of hydrogen, assuming that all the engulfed protons are captured by \nuc{12}C in the middle of convective zone.  
   This is a good approximation as far as the number ratio of engulfed protons to \nuc{12}C is sufficiently small. 
   The mole abundances of CN elements resultant from proton captures are estimated at
\begin{eqnarray}
Y_{13} / \Delta X_{\rm mix} & = & 1 - a (\Delta X_{\rm mix} / Y_{12}) \\
Y_{14} / \Delta X_{\rm mix} & = & a (\Delta X_{\rm mix} / Y_{12}) / 2 
\end{eqnarray} 
where $\Delta X_{\rm mix}$ is the amount of mixed hydrogen per unit mass and $a$ is the ratio of proton capture rates between \nuc{13}C and \nuc{12}C $[ a \equiv \langle \sigma v \rangle_{p \gamma} (\nucm{13}{C}) / \langle \sigma v \rangle_{p \gamma} ( \nucm{12}{C}) \simeq 3 \sim 4$].  
   Accordingly we may well ignore the fraction of mixed hydrogen that is captured by \nuc{13}C to form nitrogen, as long as $\Delta X_{\rm mix} / Y_{12} \ll 1$, i.e., the amount of mixed proton is much smaller than the amount of \nuc{12}{C} in the helium convection.  
   In addition, \nuc{14}N, if mixed, has little effect on our results essentially since it promptly reacts with neutron via the $\nucm{14}{N}(n, p) \nucm{14}{C}$ reaction and proton is absorbed mostly by \nuc{12}C to form \nuc{13}C. 

\begin{figure}
\FigureFile(120mm,120mm){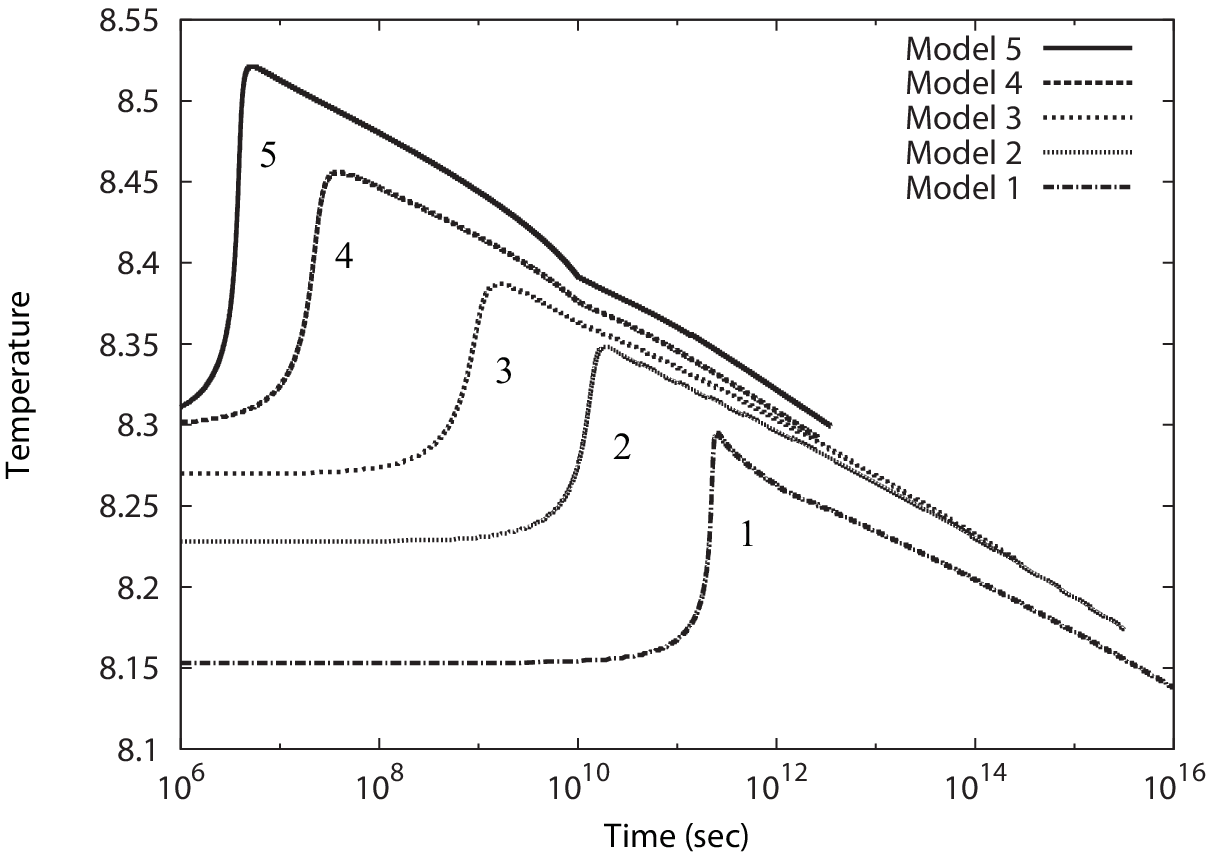}
\FigureFile(120mm,120mm){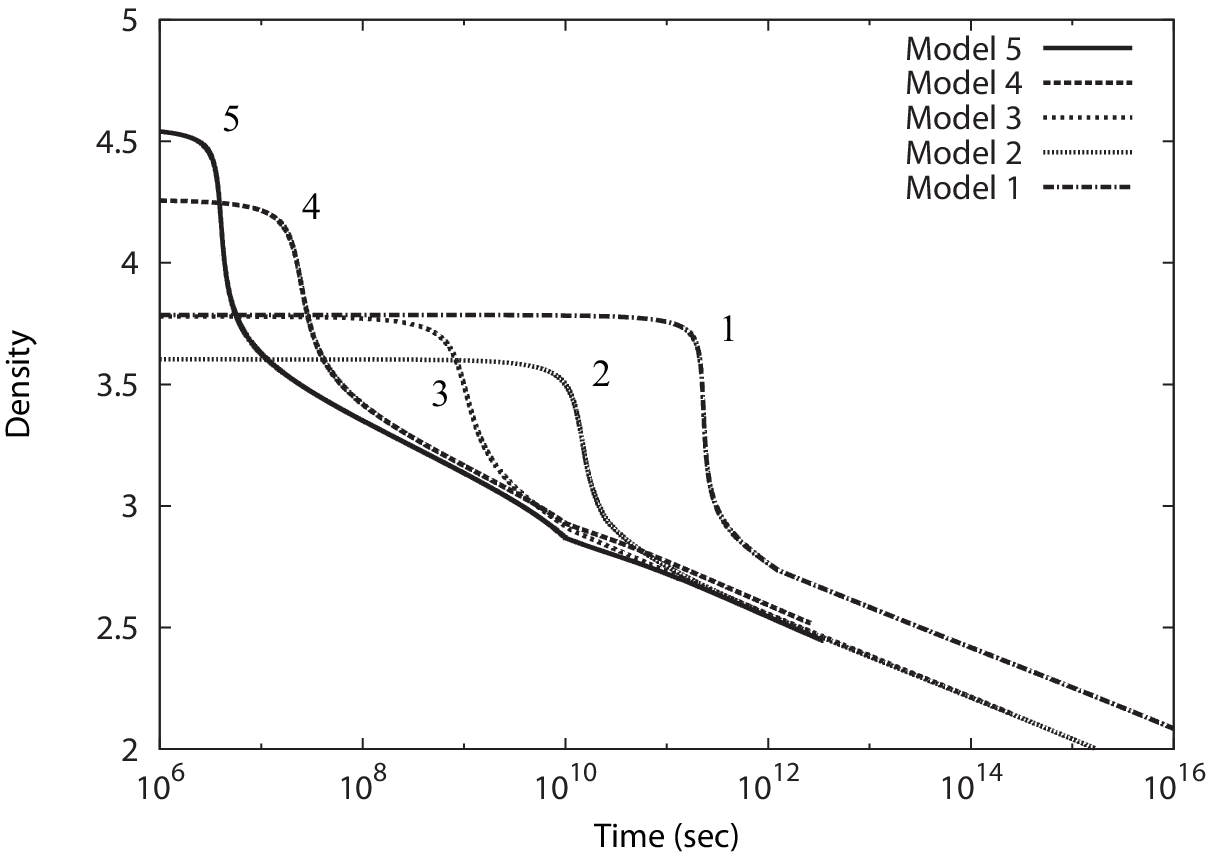}
\caption{
The structural changes during the helium shell flash, calculated with use of the analytical solution of helium flash in finite amplitude \citep[see e.g., ][]{Fujimoto1999}; 
  top and bottom panels show the time variations of temperature and density at the bottom of flash convective zone. }
\label{fig:model}
\end{figure}

The development of shell flashes is calculated with use of the analytic solution of burning shell by \citet{Sugimoto1978}, which enables the coverage of wide parameter spaces.  
We further adopt the one-zone approximation, as formulated by \citet[][see also Fujimoto et al.~1999]{Fujimoto1982b} under the assumption of the uniform distributions of abundances in the helium convective zone and with use of the averaged nuclear reaction rates over the convective zone estimated under the approximation of adiabatic gradients of density and temperature. 
  Figure~\ref{fig:model} illustrates the time variations in the temperature and density at the bottom of helium convective zone during the shell flashes. 
   The model parameters are summarized in Table~\ref{tb:para}. 
   The mass, $M_c$, and radius, $r_c$, of the core interior to the helium burning shell and the proper pressure $P_*$ at the bottom of helium burning shell are adopted from an evolutionary computations of EMP stars and metal-free Pop.~III stars when they experience the hydrogen mixing for the first time.     
   Model 1 and 3 are from the evolutionary calculation of Pop.~III models of $1.5 \msun$ and $2.0 \msun$ by \citet{Suda2008a}.  
   Model 4 is from the evolutionary calculation of $2 \msun$ and $\feoh =-3$ by \citet{Fujimoto2000}.  
   In addition, we devise Models 2 and 5 by slightly modifying Models 3 and 4, respectively, in order to cover the parameter range and in particular, the latter to see the effect of large temperature. 

   In the present study, we consider the amount, $\Delta Y_{\rm 13, mix}$, of mixed \nuc{13}C in mole per unit mass as parameter, setting $\Delta Y_{\rm 13, mix} = \Delta X_{\rm mix}$, and add it at a constant rate for an interval $\Delta t_{\rm mix}$ starting from time $t_{\rm peak}$ when the burning rate of 3$\alpha$ reaction reaches the maximum.  
   We may express the amount of mixed $^{13}$C by the mass ratio, $\dmixm$, to the $^{12}$C abundance at the start of mixing, i.e., $\dmixm \equiv \Delta X_{13, \rm mix} / X_{12} (t_{\rm peak}) = (13/12) \Delta Y_{13, \rm mix} / Y_{12} (t_{\rm peak})$.
   The amount of mixed \nuc{13}C corresponds to the amount of mixed hydrogen-rich matter as 
\begin{equation}
	\Delta M_{\rm H, mix} = 4.6 \times 10^{-6} \left( \frac{\Delta M_{\heconv}}{0.03 \msun} \right)  \left(\frac{ \dmixm}{ 0.01} \right) \left( \frac{X_{12, \heconv}}{0.2} \right) \msun, 
\label{eq:mix-hyd}
\end{equation} 
   where $\Delta M_{\heconv}$ is the mass in the helium convective zone. 
   Accordingly only a small part of the mass in the tail of hydrogen profile suffices to account for the amount of mixed hydrogen-rich matter.  
   For $\dmixm \ll 1$, this amount of mixed hydrogen is significantly smaller than reported for the stellar evolution models which develops He-FDDM \citep{Hollowell1990,Fujimoto2000,Iwamoto2004,Straniero2004}, and liable to be overlooked in the evolutionary computations. 
   As for the ignition of hydrogen shell-flash and splitting of helium flash convection, \citet{Sweigart1974} argues that it is related to the energy deposition rate due to the mixed hydrogen, and hence, to the mixing rate rather than to the amount of mixed hydrogen, and estimate $d M_{\rm H, mix} / d t \gtrsim 3 \times 10^{-14} \msun \hbox{ s}^{-1}$ for the shell flash with the peak energy generation rate due to the helium flash $L_p = 1.2 \times 10^7 L_\odot$ (and the core mass $M_c = 0.62 \msun$).  
   This is transferred to the lower limit to the duration of mixing epoch as
\begin{equation}
\Delta t_{\rm mix} > 1.6 \times 10^{8} \left( \frac{L_p}{10^7 L_\odot} \right)^{-1} \left( \frac{\dmixm}{0.01} \right) \left( \frac{\Delta M_{\rm He, conv}}{0.03 \msun} \right) \hbox{ s},
\end{equation} 
which is sufficiently smaller than the timescales of temperature decrease after the peak (see Figure~\ref{fig:model} and Table~\ref{tb:para}).

   Along the temperature and density histories of the models, we follow the progress of nucleosynthesis by adding a specific amount of \nuc{13}C to the helium convective zone in a specific interval after the peak of helium shell-flashes.  
   The initial abundances of helium, \nuc{12}C and \nuc{16}O in the helium convective zone are also taken from the models of metal-free stars.  
   As compared with them, the other metals have much smaller abundances in the helium zone of EMP stars and their effects will be discussed later based on the results.   
   The nuclear reaction network involves 59 isotopes of the 16 lightest species from the neutron and \nuc{1}H through \nuc{35}S, which is an extension of the version by \citet[][see also Aikawa et al. 2004]{Aikawa2001}.  
   Reaction rates are taken from NACRE \citep{Angulo1999} and \citet{Caughlan1988} for charged-particle reactions and \citet{Bao2000} for neutron capture reactions. 
   As for the neutron reactions of \nuc{17}O, the cross section of $\nucm{17}{O} (n, \alpha) \nucm{14}{C}$ is set at 10 mb, given by \citet{Koehler1991} in the temperature range relevant to our problems. 
   The cross section of $\nucm{17}{O} (n, \gamma)$ reactions, which is missing out of the compilation by \citet{Bao2000}, is set equal to that of \nuc{16}O.    
   The $(n, \alpha)$ reaction of \nuc{33}S and the cross sections of $(n, \gamma)$ reactions of \nuc{32}P are taken from \citet{Rauscher2000}.   

\begin{figure}
\FigureFile(120mm,120mm){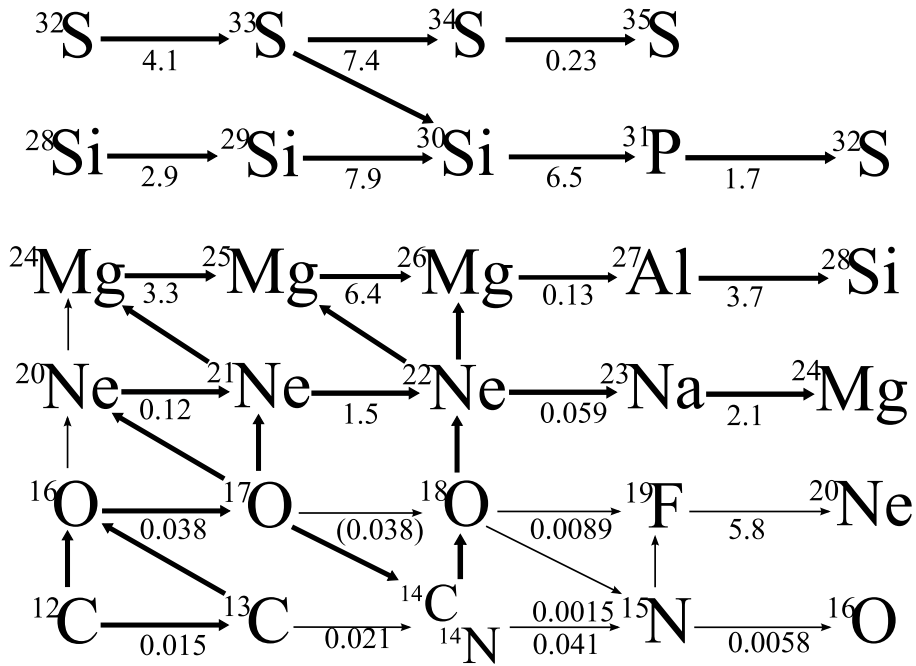}
\caption{
Flowchart of nucleosynthesis during the convective \nuc{13}C burning. 
Right arrows denote the neutron captures to which the neutron capture cross sections are attached in unit of mb at thermal energy $k T = 30$ keV, taken from \citet{Bao2000}.  
Upward arrows denote the $(\alpha, \gamma)$ reactions and upward and downward diagonal arrows denote the $(\alpha, n)$ and $(n,\alpha)$ reactions, respectively.  
   Thick arrows represent the main paths of nucleosynthesis.  
}
\label{fig:chart}
\end{figure}

Figure~\ref{fig:chart} outlines the main paths of nucleosynthesis from \nuc{12}C through \nuc{35}S via the $\alpha$ and neutron captures. 
   Under the presence of neutron source, the $(\alpha, n)$ reactions coexist with the $(\alpha, \gamma)$ reactions.  
   In particular, the $\nucm{A}{Z} (n, \gamma) \nucm{A+1}{Z} (\alpha, n) \nucm{A+4}{(Z+2)}$ reaction may compete with the $\nucm{A}{Z} (\alpha, \gamma) \nucm{A+4}{(Z+2)}$ reaction to form $\alpha$-nuclei of mass number $A+4$ from $\alpha$-nuclei of mass number $A$.  
   This competition may be expressed in terms of the ratio, $\lambda_{\rm A+4}$, of production rates of $\nucm{A+4}{(Z+2)}$ nuclei at the moment $t$, defined as: 
\begin{equation}
\lambda_{\rm A+4}(t) = \frac{Y_{A+1} (t) Y_{\alpha} (t) \rho(t) N_{A} \langle \sigma v \rangle_{\alpha n} (\nucm{A+1}{Z}) } {Y_{A} (t) Y_{\alpha} (t) \rho(t) N_{A} \langle \sigma v \rangle_{\alpha \gamma}( \nucm{A}{Z}) }, \nonumber
\label{eq:lambda}
\end{equation} 
   which represents the dependence of the production of the isotopes on the reaction paths.  
   Here, $Y_{A} (t)$, $\rho(t)$ and $N_{A} \langle \sigma v \rangle_{a b}$ are the number abundance in moles, the density and the rate of ($a, b$) reaction, respectively. 
   To investigate the influence of neutrons, $Y_{A+1} (t)$ is assumed to be created only through $\nucm{ A}{Z} (n, \gamma) \nucm{ A+1}{Z}$ reaction without the destruction by the $(\alpha, n)$ reaction. 
   We denote the production rate ratio under this assumption by $\lambda^*$.  
   For simplicity, we assume
\begin{eqnarray}
Y_{n} (t') &=& 0 \qquad (t' < t_{\rm mix})  \nonumber\\
{ Y_{A} (t')} /{Y_{A} (t)} &\sim& 1 \qquad (t' > t_{\rm mix}) \\
\langle \sigma v \rangle_{n \gamma} (\nucm{A}{Z}) &\sim& \sigma_{n\gamma}(\nucm{A}{Z})  v_n \nonumber
\end{eqnarray}
where $t_{\rm mix}$ means the moment when the mixing starts.  
   The first and second equations assume that there are neither neutrons nor $\nucm{A+1}{Z}$ before the mixing and little changes in most of $\alpha$-elements after the mixing, respectively.  
   In the third equation, the cross section and the thermal velocity, $v_n$, of neutrons are assumed to be constant instead of $\langle \sigma v \rangle_{n \gamma}$.  
   Then we can get the simple equation to estimate the effect of neutrons, 
\begin{eqnarray}
\lambda^*_{A+4} (t)  &=&  \tau(t) \sigma_{n \gamma} (\nucm{A}{Z}) \frac{N_{A} \langle \sigma v \rangle_{\alpha n} (\nucm{A+1}{Z})} {N_{A} \langle \sigma v \rangle_{\alpha \gamma}(\nucm{A}{Z})}, 
\label{eq:lambdam}
\end{eqnarray}
where the exposure $\tau(t)$ is defined as 
\begin{equation}
\tau(t) = \int_{0}^{t} {Y}_{n} (t') \rho(t') { N_{A}} v_n dt'.  
\end{equation}
   The above assumptions will not strongly affect our discussion because in almost all the cases $\lambda^*_{A+4} (t)$ results in such large values as vary by several orders of magnitude (see below). 
 
Figure~\ref{fig:lifetimes} shows the comparison of the $(\alpha, \gamma)$ and $(\alpha, n)$ reaction rates for the light elements.  
   The $\nucm{A+1}{Z} (\alpha, n) \nucm{A+4}{Z} $ rates are much larger than the $\nucm{A}{Z} (\alpha, \gamma) \nucm{A+4}{Z} $ rates.
   Their differences amount to many orders of magnitude for carbon and oxygen.    
   Thus, $\lambda^*_{A+4} (t) \gg 1$ so that the $(n, \gamma)$ and $(\alpha, n)$ reactions dominate the formation of $\alpha$-nuclei even under small neutron exposures. 
   It should be noted, however, that the actual ratio of the formation rates results much smaller than $\lambda^*_{A+4} (t)$ if we take into account the destruction of ${}^{A+1}Z$ nuclei by $\alpha$-capture [$\lambda_{A+4} (t) \simeq (d \tau / d t) \sigma_{n\gamma} (\nucm{A}{Z}) / \rho (t) N_{A} \langle \sigma v \rangle_{\alpha \gamma}( \nucm{A}{Z}) $].  
    For magnesium and beyond, the $\alpha$-capture reactions themselves may reduce to be too slow to occur during the helium shell flashes.  
    For heavier nuclei, on the other hand, the cross sections of neutron capture grow larger in general.  
	Accordingly, with under a small neutron exposure, the helium burning can drive the nucleosynthesis up to heavier nuclei than otherwise expected, via the combinations of neutron capture and $(\alpha, n)$ reactions.  

\section{Results}

We compute the progress of nucleosynthesis in the helium convective zone during the helium shell flashes in the metal-free stars, born of the gas totally lacking metals, by adding \nuc{13}C for the thermal history given by Models 1-5 in Figure~\ref{fig:model}.  
   The progress of nucleosynthesis is exemplified in Figures~\ref{fig:synthesis}, where the variations of the abundances are plotted for the major elements and isotopes against the elapsed time from the onset of shell flash (left panel) and against the neutron exposure (right panel) for Model 4 with $\dmixm = 10^{-2}$ and $\Delta t_{\rm mix} = 10^9$ s.  
   As soon as the mixing starts, the $\nucm{13}{C} (\alpha, n) \nucm{16}{O}$ reaction proceeds in much shorter timescale than 3$\alpha$ reaction and the neutron density soon reaches as large as $N_n \simeq 10^{11} \hbox{ cm}^{-3}$. 
   Because of the lack of pristine metals as absorbers, neutrons are exclusively react with the most abundant nucleus \nuc{12}C and enter into the neutron-recycling reactions of $\nucm{12}{C} (n, \gamma) \nucm{13}{C} (\alpha, n) \nucm{16}{O}$ \citep{Gallino1988}.  
   Along with the run of neutron-recycling reactions, the second most abundant nucleus \nuc{16}O increases in number, eventually to absorb a neutron and to cause the leak from the recycling circuit.  

\begin{figure}
\FigureFile(120mm,120mm){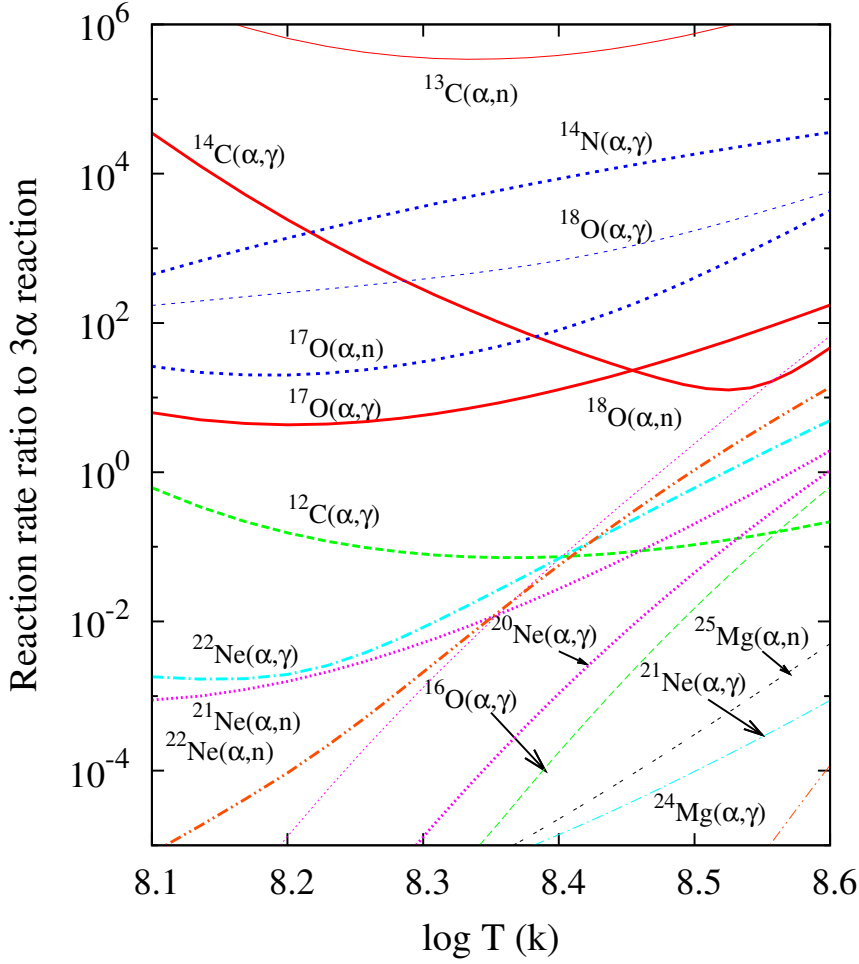}
\caption{
The rates of $\alpha$-capture reactions $d \log Y_i / dt = Y_4 \rho N_A \langle \sigma v \rangle_{\alpha \gamma}$ and $Y_4 \rho N_A \langle \sigma v \rangle_{\alpha n}$, relative to the 3$\alpha$ reaction rate ($d \log Y_4/dt = 3Y_4^{2} \rho^2 N_A^2 \langle \sigma v \rangle_{\alpha\alpha\gamma}/6$), as a function of temperature. 
   Rates are taken from \citet{Angulo1999} except for the rates of $\nucm{14}{C} (\alpha, \gamma) \nucm{18}{O}$, $\nucm{16}{O} (\alpha, \gamma) \nucm{20}{Ne}$, $\nucm{17}{O} (\alpha, \gamma) \nucm{21}{Ne}$, $\nucm{21}{Ne} (\alpha, \gamma) \nucm{26}{Mg}$, $\nucm{24}{Mg} (\alpha, \gamma) \nucm{28}{Si}$ and $\nucm{25}{Mg} (\alpha, n) \nucm{28}{Si}$ taken from \cite{Caughlan1988}. 
   We adopt 3$\alpha$ reaction rate from \cite{Caughlan1988} for the $\alpha$-capture reaction rates, enumerated above.
   The density and helium abundance are set at $\rho = 10^3 \hbox{ g cm}^{-3}$ and $Y_4 = 0.2$, respectively.  
}
\label{fig:lifetimes}
\end{figure}

   The formation of \nuc{17}O opens various channels of reactions, consisting of the neutron and $\alpha$-captures.   
   Because of much larger cross section, \nuc{17}O soon captures another neutron to be \nuc{14}{C}, and the latter is converted into \nuc{18}O and then into \nuc{22}Ne by successive $\alpha$-captures.  
   The abundances of \nuc{17}{O}, \nuc{14}{C}, and \nuc{18}{O} increase as \nuc{13}C is added, and yet, remain much smaller than the amount, $\Delta Y_{\rm 13, mix}$, of the mixed \nuc{13}C.  
In this case, \nuc{14}C attains to a fairly large abundance of $\sim 0.2 \Delta Y_{\rm 13, mix}$ since the rate of $\alpha$-capture decreases relative to 3$\alpha$-reaction rates with increasing temperature, as seen from Figure~\ref{fig:lifetimes}.  
   At the same time, \nuc{17}{O} reacts via the $(\alpha, n)$ channel to produce \nuc{20}Ne and releases neutron.  
   This reaction is slower by a factor of $\sim 10^{4}$ than the $\nucm{13}{C} (\alpha, n) \nucm{16}{O}$ reaction, but, sufficiently faster than the 3$\alpha$ reactions. 
   The released neutrons are mostly captured by \nuc{12}C and enter into the doubly neutron-recycling reactions of $\nucm{12}{C} (n, \gamma) \nucm{13}{C} (\alpha, n) \nucm{16}{O} (n, \gamma) \nucm{17}{O} (\alpha, n) \nucm{20}{Ne}$.  
   In addition to the neutron capture of \nuc{17}O, this cycle has a leakage of the $\nucm{17}{O}(\alpha, \gamma)$ channel to \nuc{21}{Ne}, and it captures neutron to be \nuc{22}Ne with a fairly large cross section.   

The two paths from \nuc{17}O meet at \nuc{22}Ne.  
The latter has a small cross section of neutron capture, comparable with \nuc{12}C and \nuc{16}O, and the $\alpha$-capture rates are slower than the 3$\alpha$ reactions for the temperature of $T_{\rm max} \lesssim 3 \times 10^8$ K.  
   Accordingly, the flow of nucleosynthesis is bottlenecked and neutrons stagnate in \nuc{22}Ne.  
   As a corollary, the abundance of \nuc{22}Ne reaches up to a half amount of the mixed \nuc{13}C and $Y_{22} \simeq \Delta Y_{\rm 13, mix} / 2$ since two neutrons are spent to form \nuc{22}Ne.  

   As neon isotopes accumulate, sodium and Mg isotopes are created, though in much smaller abundances than \nuc{22}Ne.  
   There forms a steep drop in the abundances between \nuc{22}Ne and \nuc{23}Na with the ratio $Y_{23} / Y_{22} \le \sigma_{n \gamma} (\nucm{22}{Ne}) / \sigma_{n \gamma} (\nucm{23}{Na}) = 0.028$, where the equality holds for a large neutron exposure of $\tau \gg 1/ \sigma_{n \gamma} (\nucm{23}{Na}) = 0.48 \hbox{ mb}^{-1}$.   
   Among Mg isotopes, the heaviest isotope \nuc{Mg}{26} tends to have larger abundance than \nuc{23}{Na} because of much smaller neutron capture cross section.
   The $\alpha$-captures of Ne isotopes may produce the Mg isotopes directly by skipping \nuc{23}Na, though they work only partially at the temperature of this model. 
   This also results in the largest abundance of \nuc{26}Mg because of larger rates and larger abundance of \nuc{22}Ne than those of \nuc{20}Ne and \nuc{21}Ne.

   The elements heavier than Mg are formed solely by neutron captures (see Figure~\ref{fig:lifetimes}).    
   They are produced late and the resultant abundances are much smaller than those of Na and Mg isotopes. 
   Because of very small neutron capture cross section of \nuc{26}Mg, there forms another sharp drop of the abundances between \nuc{26}Mg and \nuc{27}Al by more than $Y_{27} / Y_{26} < \sigma_{n \gamma} (\nucm{26}{Mg}) / \sigma_{n \gamma} (\nucm{27}{Al}) = 0.034$.  
   Among silicon isotope, \nuc{28}Si reaches as abundant as \nuc{27}Al and so does \nuc{31}P, reflecting their neutron capture cross sections much smaller than those of the ambient nuclei.  
   Among the sulfur isotopes, \nuc{33}S is depressed because of the turn-around cycle via $\nucm{33}{S} (n,\alpha) \nucm{30}{Si}$ reaction. 
   In comparison, \nuc{34}S has much larger abundance because of small neutron capture cross section.  
 
As soon as the mixing of \nuc{13}C ceases, the neutron density drops sharply to retard the neutron capture reactions, while the $\alpha$-captures of the remainders of neutron-rich C and O isotopes keep adding the abundances of neon isotopes.  
In Models 4 and 5 of relatively high temperatures, neon isotopes also burn to form Mg isotopes.  
  Accordingly, neutron supply may continue by $\nucm{17}{O} (\alpha, n) \nucm{20}{Ne}$ and $\nucm{22}{Ne} (\alpha, n) \nucm{25}{Mg}$, although the neutron density is by far smaller and further decreases with the drop of temperature.  
 
\begin{figure}
\FigureFile(120mm,120mm){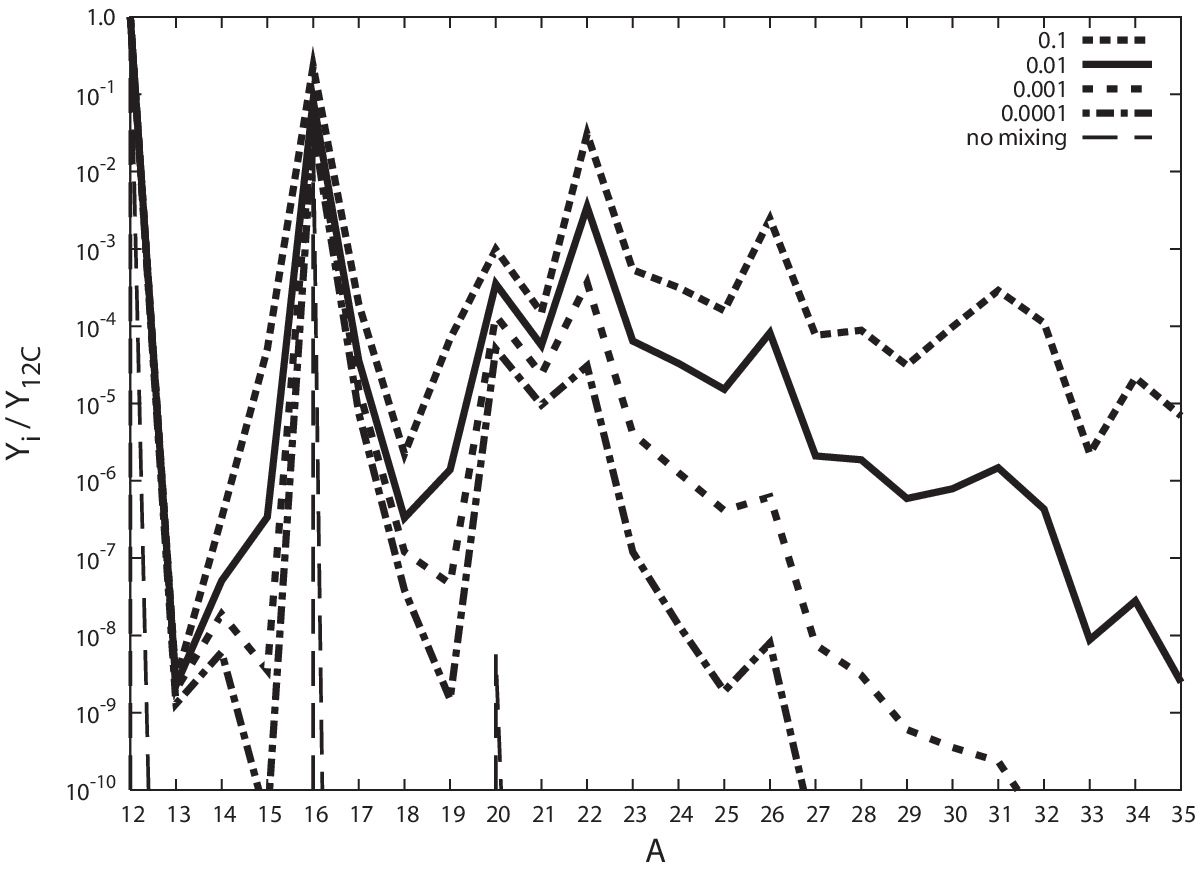}
\FigureFile(120mm,120mm){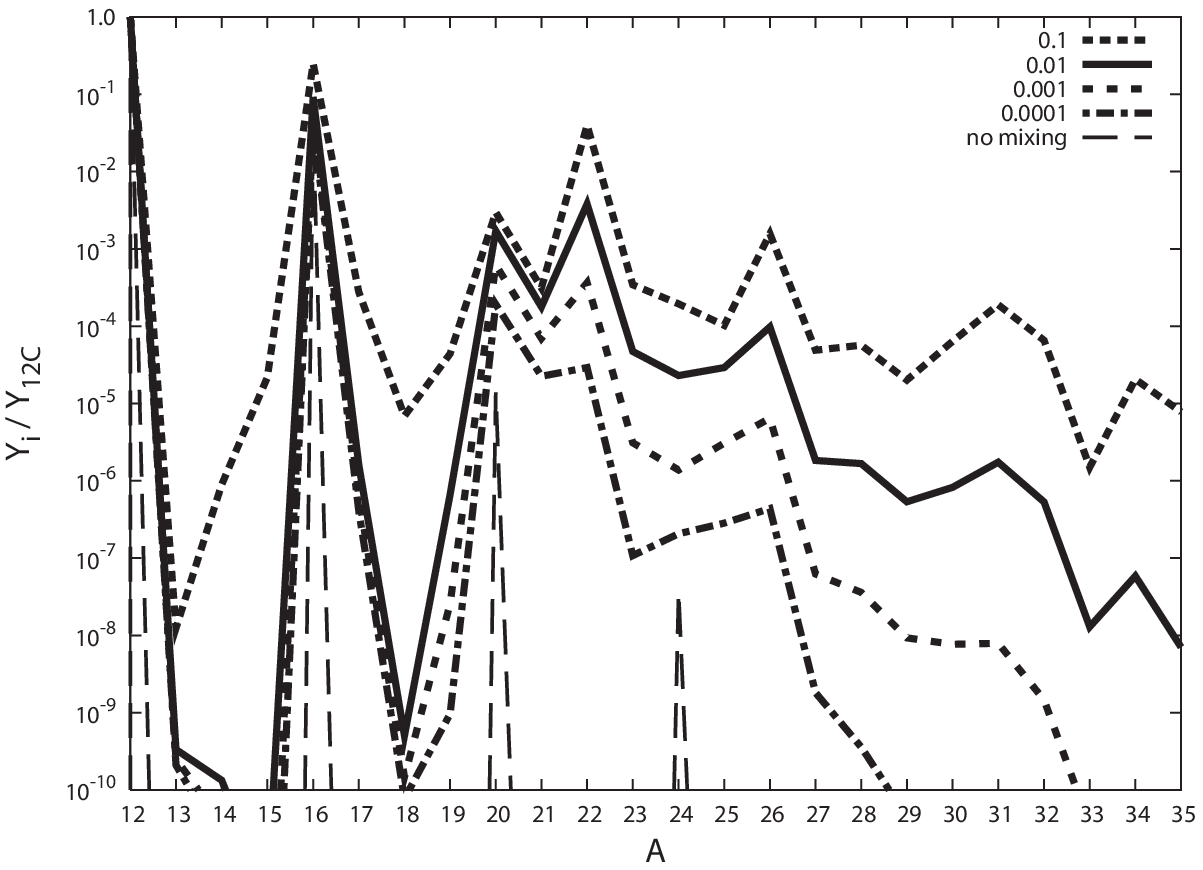}
\caption{
The number abundances of nuclear products, normalized by carbon abundance, are plotted as a function of mass number, $A$, resulting from the nucleosynthesis in the helium shell flash convective zone. 
   The results are shown for Model 1 with $\Delta t_{\rm mix} = 10^{12}$ s (top panel) and Model 5 with $\Delta t_{\rm mix} = 10^{9}$ s (bottom), respectively.
   Each line denotes the yield for the different amount of mixed \nuc{13}C, $\dmixm$ as indicated in top-right corner of each panel. 
}
\label{fig:yields}
\end{figure}

  Figure~\ref{fig:yields} shows the dependence of yields on the degree of mixing for two cases of Model 1 (top panel) and Model 5 (bottom panel), with the lowest and highest temperatures, respectively.
   The hydrogen mixing makes striking difference in the variety and amount of synthesized elements when compared with the yields of the helium shell-flashes without the \nuc{13}C mixing, which is also plotted in this figure.  
   Without the hydrogen mixing only the $(\alpha, \gamma)$ reactions produce \nuc{16}O, and also, traces of \nuc{20}Ne and \nuc{24}Mg for the flashes of high temperature.  
   In contrast, the hydrogen mixing enables to produce not only the heavier $\alpha$ nuclei in much larger amounts, but also, their neutron-rich isotopes and the odd nuclei appreciably.  

   The amounts of yields increase with the amount of mixed \nuc{13}C. 
   There is a general tendency that the dependences grow stronger for heavier elements, though varying with elements. 
   Among the oxygen isotopes, \nuc{16}O is produced via the recycling reactions, and the resultant abundance depends rather weakly on the mixed amount of \nuc{13}C. 
   There is little difference among the models.     
   On the other hand, the heavier isotopes differ in their abundances greatly between Model 1 and Model 5 since they are mostly destroyed by $\alpha$-captures after the cease of mixing.
   The latter is also the case for nitrogen isotopes. 
   In contrast, the fluorine depends weakly on the models and increases largely with the mixing amount of \nuc{13}C, which may result from the competition of two formation paths (see below). 

   Among the neon isotopes, the heaviest isotope \nuc{22}Ne increases nearly in proportion to the amount of mixed \nuc{13}C, as stated above. 
   In comparison, the lighter isotopes, \nuc{20}Ne and \nuc{21}Ne, have weaker dependence on the amount of mixed \nuc{13}C as a result of the competition between the $\alpha$ and neutron captures of \nuc{17}O;  
   for a smaller mixing of \nuc{13}C, the $\alpha$-capture dominates over the neutron capture to produce more \nuc{20}Ne than \nuc{22}Ne and vice versa for a larger mixing.  
   The temperature dependences are discernible in the larger abundances of \nuc{20}Ne for Model 5 and in the survival of neutron-rich carbon and oxygen isotopes and of nitrogen in larger amounts for Model 1.    

   Beyond neon, the dependence on the amount of mixed \nuc{13}C grows steeper with the mass number while the relative distribution abundances approach to the steady-state distributions of neutron captures for larger mixing.  
   Among the magnesium isotopes, the heaviest isotope \nuc{26}Mg has the largest abundance, and the influence of $\alpha$-captures is discernible for the shell flashes of the higher peak temperatures.  
   For still heavier elements, the dependence remains very strong even for the largest mixing. 
  The abundances of phosphorus and sulfur isotopes vary by more than a third power of the amount of mixed \nuc{13}C.  

\begin{figure}
\FigureFile(120mm,120mm){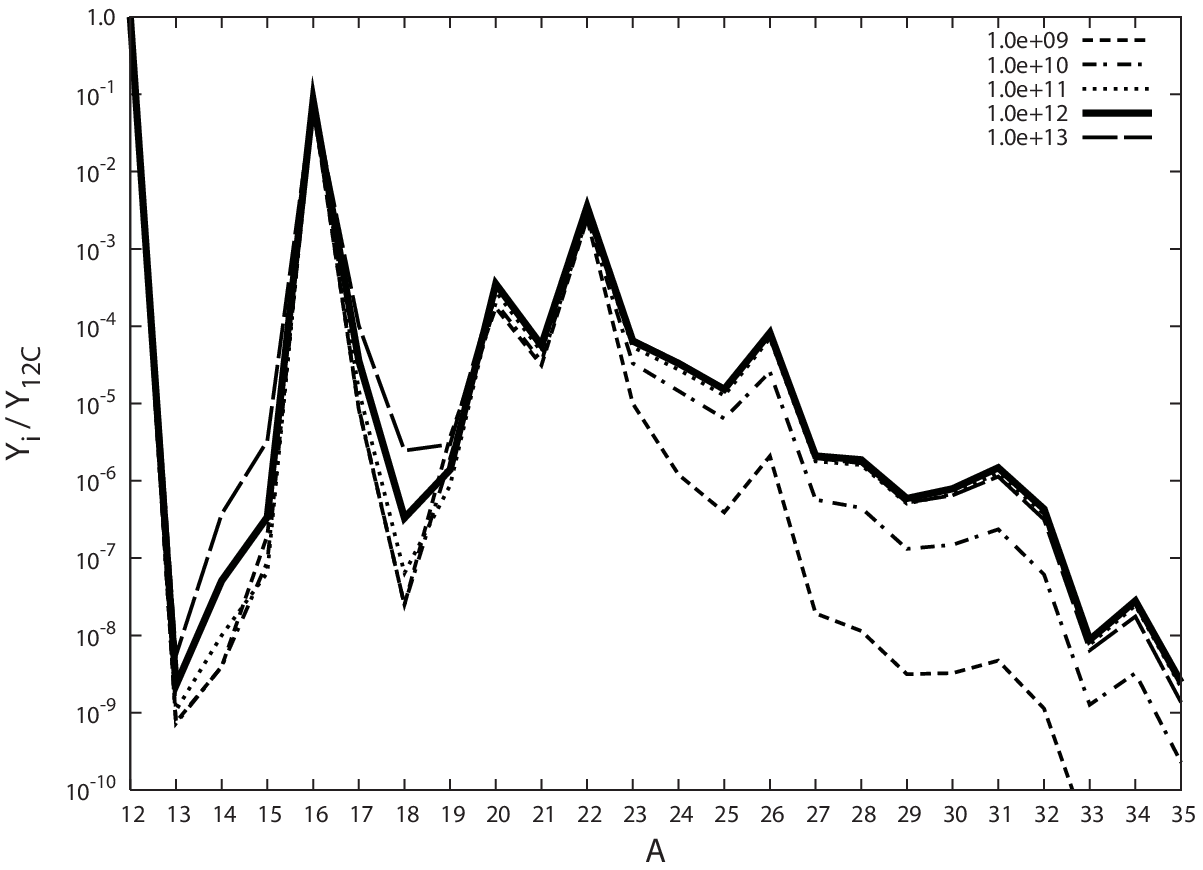}
\FigureFile(120mm,120mm){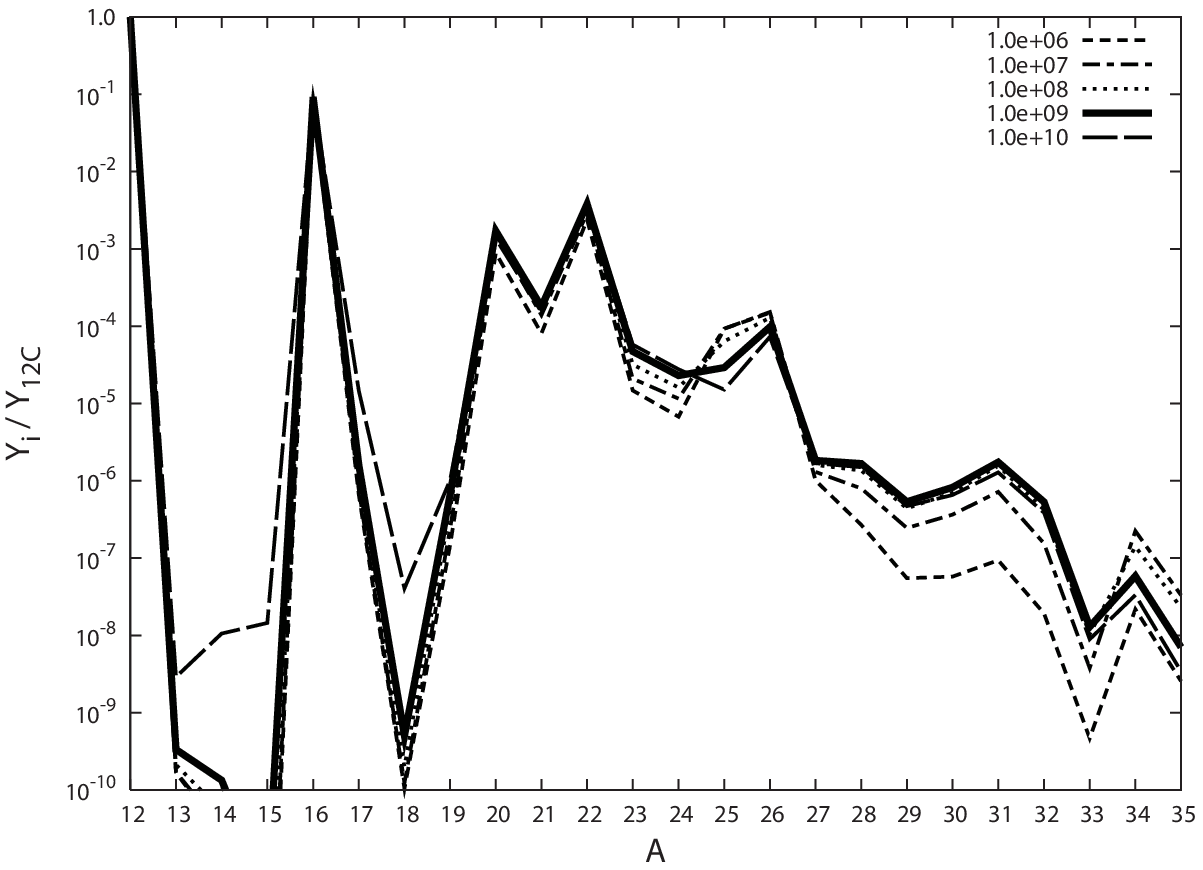}
\caption{
The same as in Figure~\ref{fig:yields} but for the different durations, $\Delta t_{\rm mix}$, of mixing epoch with the same amount of mixed \nuc{13}C, $\dmixm=0.01$ for Models 1 (top panel) and 5 (bottom panel):   
   numerals in the top-right corner of each panel denote the durations of mixing epochs in units of second.
}
\label{fig:mixtime-dep}
\end{figure}

Figure~\ref{fig:mixtime-dep} compares the yield abundances, resulting from different mixing durations for the same amount of mixed \nuc{13}C.  
The abundance patterns are almost independent of mixing duration for sufficiently long mixing epoch, while the production of heavier nuclei is suppressed for the shorter durations of $\Delta t \lesssim 10^7$ s for Model 5 and $10^{10}$ s for Model 1. 
These durations are related to the timescales necessary for $\alpha$-captures to synthesize \nuc{22}Ne that acts as seed for the synthesis of heavier nuclei.  
   This fact manifests itself in the smaller abundance of \nuc{20}Ne for the short mixing durations.   
   Further, the mixing duration also has an effect on the $\alpha$ -captures after the end of \nuc{13}C mixing.  
   In Model 5, \nuc{25}Mg and \nuc{26}Mg are more abundant for shorter durations since they are formed via $\alpha$-captures of \nuc{22}Ne after the neutron irradiation has ceased.   
   In Model 1, we see that \nuc{14}C or \nuc{14}N, \nuc{15}N, \nuc{17}O and \nuc{18}O remain with larger abundances for longer durations, which is simply because of shortage of time and temperature to burn them via $\alpha$-capture after the cease of mixing.    

\begin{figure}
\FigureFile(120mm,120mm){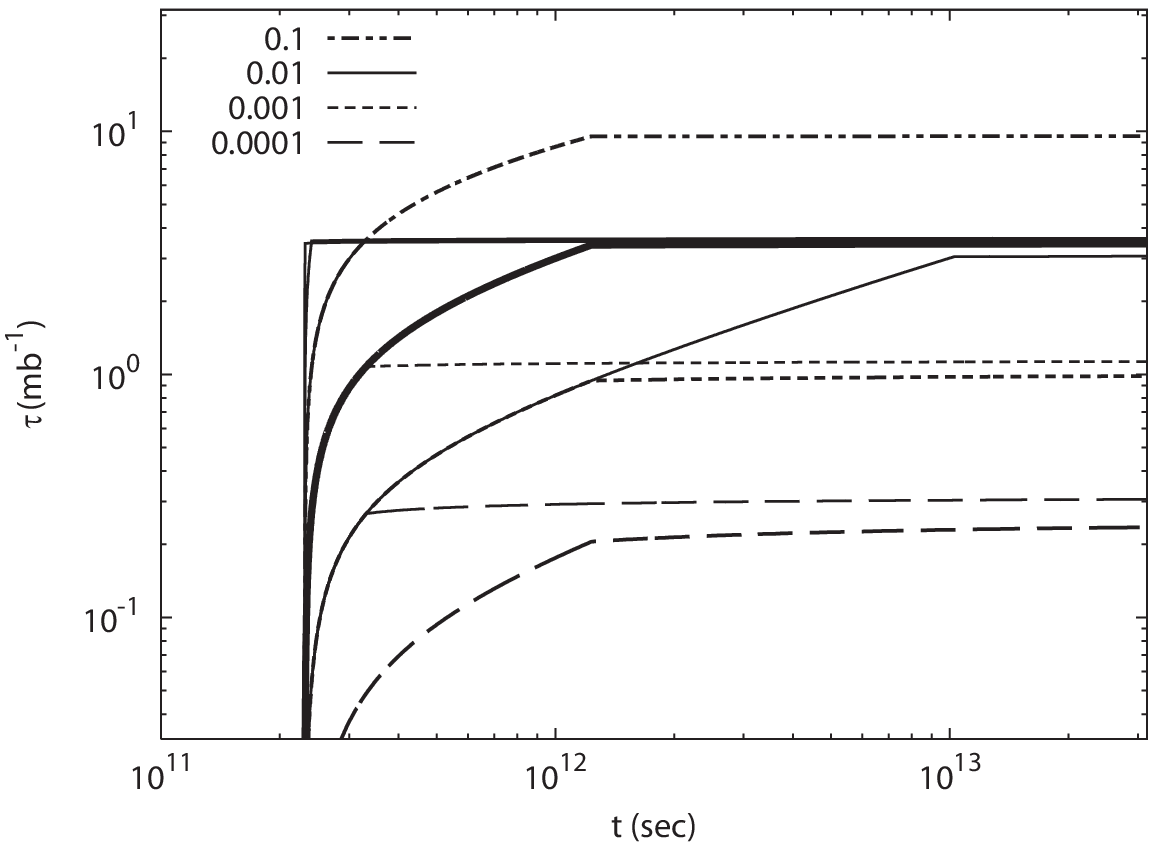}
\FigureFile(120mm,120mm){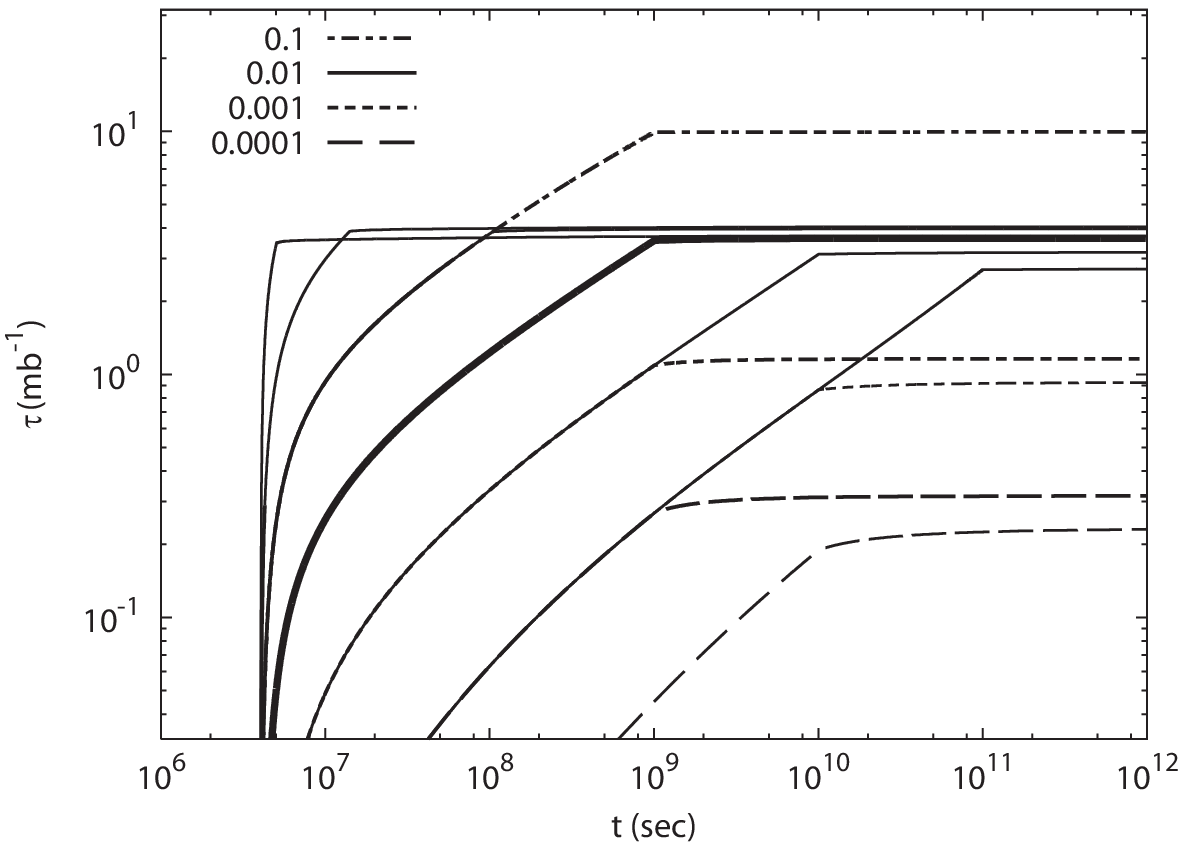}
\caption{
The neutron exposure as a function of time for Model 1 (top panel) and Model 5 (bottom panel) with the different amounts and duration of \nuc{13}C mixing.  
   The amounts of mixed \nuc{13}C, $\dmixm$, are denoted by different lines as indicated in the top-left corner. 
   The durations of mixing $\Delta t_{\rm mix}$ are read from a break in each curve.
}
\label{fig:n-exp}
\end{figure}

Figure~\ref{fig:n-exp} shows the variations of neutron exposure during the shell flashes.  
   It is determined essentially by the amount of mixed \nuc{13}{C} and increases nearly in proportion to its square root because of concomitant production of neutron absorbers. 
   The weak dependence on the mixing duration results from the competition between the $\alpha$ and neutron captures of \nuc{17}O.  
   For the shorter durations, large neutron density suppresses the second-recycling reaction while for the longer duration, the production of heavier elements increases the average cross sections of neutron capture, both of which work to diminish the neutron exposure.

In summary, in the nucleosynthesis during the helium flashes with the hydrogen mixing, neutrons play a critical role in accelerating $\alpha$-captures to create the $\alpha$ nuclei of oxygen through magnesium as catalytic agents, and also, to generate their neutron-rich isotopes and odd nuclei as incorporated.  
   In the following, we discuss the formation paths to the light $\alpha$-nuclei, their neutron-rich isotopes and related odd nuclei separately.

\subsection{Formation Path to Particular Elements} 
  
\subsubsection{Oxygen Isotopes and Fluorine} 

\begin{figure}
\FigureFile(120mm,120mm){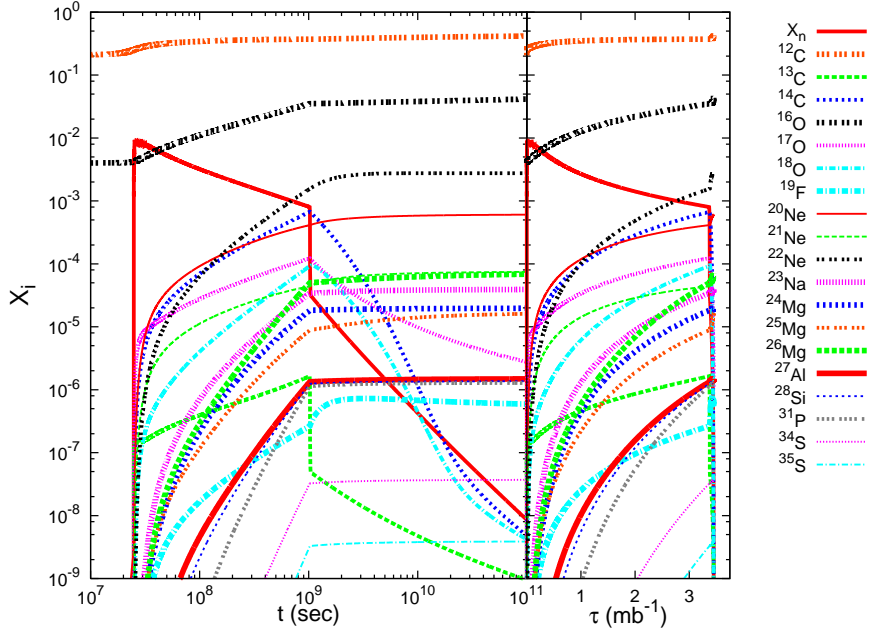}
\caption{
The progress of nucleosynthesis, induced by hydrogen mixing, in the helium-flash convection.   
   The elemental abundances are plotted against time (left panel) and against the neutron exposure, $\tau$ (right panel), respectively;  
   the neutron abundance, $X_n$, is multiplied by $10^{13}$.  
   The results are taken from Model 4 with the amount of mixed \nuc{13}C, $\Delta X_{\rm 13, mix} = 0.01 X_{12} (t_{\rm peak})$ in total, (where $X_{12} (t_{\rm peak})$ is the abundance of \nuc{12}C in the helium convective zone at the peak of flash, $t= t_{\rm peak}$), or $\dmixm = 0.01$, which is supplied at a constant rate for an interval of $\Delta t_{\rm mix} = 10^9$ s starting from the peak of shell flash. 
}
\label{fig:synthesis}
\end{figure}

To create $^{16}$O, there are two paths, $(\alpha, \gamma)$ and $(n, \gamma)$ plus $(\alpha, n)$, from \nuc{12}C in the helium zone with the hydrogen mixing. 
The $\nucm{13}{C} (\alpha, n) \nucm{16}{O}$ reaction is faster by many orders of magnitude than the $\nucm{12}{C} (\alpha, \gamma) \nucm{16}{O}$ (see Figure~\ref{fig:lifetimes}), and the ratio of production rates is estimated at $\lambda^*_{^{16}{\rm O}} \simeq 10^{5}$ for the Model 4 of $\dmixm = 10^{-2}$ in Fig~\ref{fig:synthesis}.   
   Accordingly, the ($n, \gamma$) and ($\alpha, n$) reactions are the main channel to \nuc{16}O at least while the \nuc{13}C mixing continues.  

The amount of \nuc{16}O, thus produced, results from the competition between the neutron capture by \nuc{12}C and \nuc{16}O.  
   Neutrons, if captured by \nuc{12}C, enter into the neutron-recycling reaction $\nucm{12}{C} (n, \gamma) \nucm{13}{C} (\alpha, n) \nucm{16}{O}$.
   On the other hand, if absorbed by \nuc{16}O, they escape from the recycling circuit and finally end by settling in the neutron-rich isotopes from \nuc{14}C up to \nuc{22}{Ne}.  
   By neglecting a small variation of \nuc{12}C abundance, we may evaluate this competition analytically (see Appendix 1).   
   For $\tau \ll 1 / \sigma_{n \gamma} (\nucm{16}{O}) = 26 \hbox{ mb}^{-1}$, the resultant oxygen abundance is given as a function of the amount of mixed \nuc{13}C from eq.~(\ref{eq:o-proa14}) by 
\begin{equation}
	Y_{16} / Y_{12} \simeq  \left[ {2(1 + \xi) \over 2 + \xi(1 + f) } {\sigma_{n \gamma} (\nucm{12}{C}) \over \sigma_{n \gamma} (\nucm{16}{O})} \right]^{1/2} Y_{12}^{1/2}  (\Delta Y_{\rm 13, mix} / Y_{12})^{1/2},  
\label{eq:ox-pro}
\end{equation}
   where $\xi$ denotes the ratio of reaction rates between the $\alpha$ to neutron captures of \nuc{17}O and $f$ denotes the branching ratio of the $(\alpha, \gamma)$ to the total reactions with $\alpha$ particles.  
  For still larger $\tau$, an upper bound is set at a steady state value of
\begin{equation}
Y_{16} / Y_{12} <  \sigma_{n \gamma} (\nucm{12}{C}) / \sigma_{n \gamma} (\nucm{16}{O}) = 0.40. 
\label{eq:ox-pro2}  
\end{equation}

On the other hand, the $\nucm{12}{C} (\alpha, \gamma) \nucm{16}{O}$ reaction works in longer time, producing an amount of $\delta Y_{16} = 2.9 \times 10^{-3}$ in total and $\delta Y_{16} = 2.19 \times 10^{-3}$ after the mixing has ceased for Model 4 in Figure~\ref{fig:synthesis}. 
   As the convection extends outwards during the helium flashes, carbon and oxygen suffer from dilution with helium in the helium zone newly incorporated.  
   Consequently, as the shell flashes recur without hydrogen mixing, the ratio of carbon to oxygen approaches their production rates, $\dot{\rm C}$ and $\dot{\rm O}$, due to the $3\alpha$ and $\nucm{12}{C}(\alpha, \gamma)\nucm{16}{O}$ reactions, respectively \citep{Suda2004}.  
   Thus, we have 
\begin{equation}
Y_{16} / Y_{12} \simeq \dot{\rm O} / \dot{\rm C} \simeq (0.07 \mathchar`- 0.4) 0.25 (X_{12 } /0.2) (0.8 / Y)^ 2 (10^3 \hbox{ g cm}^{-3} / \rho) 
\label{eq:ox-proag}
\end{equation} 
   for the relevant temperature range ($\log T =8.2 \sim 8.5$) where $Y$ and $X_{12}$ are the helium and carbon abundances in the helium convective zone [$Y + X_{12} (1 +  \dot{\rm O} / 0.75 \dot{\rm C}) = 1$].  
   This sets the range of oxygen abundance resulting from the third dredge-up only. 

Other isotopes, \nuc{17}O and \nuc{18}O, are produced by the $\nucm{16}{O} (n, \gamma) \nucm{17}{O}$ reaction and by the subsequent reactions of $\nucm{17}{O} (n, \alpha) \nucm{14}{C} (\alpha, \gamma) \nucm{18}{O}$, respectively.  
Because of much larger cross section of neutron capture, \nuc{17}O remains much smaller in the abundance than \nuc{16}O [$Y_{17}/Y_{16} \le 0.0038/(1 + \xi)$] while irradiated by neutrons. 
After the cease of \nuc{13}C mixing, it is effectively destroyed by the $\alpha$-captures. 
   The same is true for \nuc{14}C (or its daughter nuclei \nuc{14}N) and \nuc{18}O.
   They accumulate as \nuc{13}C is added, and decreases rapidly after the neutron supply stops.
   The abundance of \nuc{18}O finally reduces much smaller than that of \nuc{17}O, which reflects much larger $\alpha$-capture rates of \nuc{18}O.  

As \nuc{18}O increases, \nuc{19}F is formed, though much smaller in the abundance.  
   Under neutron-rich circumstance, there are two channels of $\nucm{18}{O} (n, \gamma) \nucm{19}{O} (e^{-}\bar{\nu}) \nucm{19}{F}$ and $\nucm{15}{N}(\alpha, \gamma) \nucm{19}{F}$ \citep[see][]{Mowlavi1998}.  
   The latter $\alpha$-capture channel dominates the former channel even for large \nuc{13}C mixing; 
   the latter contribution accounts for $2/3 \sim 0.9$.  
   This can be discernible in Figure~\ref{fig:synthesis}, where the abundance is a few times larger than expected from that of \nuc{18}{O} under the steady-state of neutron capture during the mixing and increases rapidly via $\alpha$-capture of \nuc{15}N after the cease of mixing and the resultant drop of neutron density. 
   Although the abundance is itself not large, this can give rise to large enrichments because of its smaller solar abundance, reaching the enrichment comparable with carbon for the largest mixing.  

\subsubsection{Neon isotopes and Sodium}

The argument similar to the \nuc{16}O case is applicable to the effective paths to create \nuc{20}{Ne}. 
Since $\lambda^*_{\nucm{20}{Ne}} \sim 10^4$ in Model 4 of Fig~\ref{fig:synthesis}, the main path is expected to be the $(n,\gamma)$ and $(\alpha, n)$ reactions as well.  
Differently from \nuc{16}O case, however, the $\alpha$-capture of \nuc{17}O has the $(\alpha, \gamma)$ branch, which works as effective neutron leakage out of doubly neutron-recycling circuit.  
   Thus, the abundance ratio is determined by the branching ratio as $Y_{21} / Y_{20} \simeq 1/5.5 -1/10$ around the temperatures ($T_8 \simeq 2 -3$) of interest here \citep{Caughlan1988,Angulo1999}.   
   In addition, the formed \nuc{21}Ne may absorb neutron while \nuc{20}Ne largely survives for the large neutron exposure.  
   As neutron capture proceeds, the abundance ratio decreases and approaches the steady-state values of $Y_{21} / Y_{20} = 0.08 /(1 - f)$, which happens to be only slightly smaller than given by the branching ratio.   
   It also decreases slightly at high temperatures because of larger $\nucm{21}{Ne} (\alpha, n)$ rate than $\nucm{20}{Ne}(\alpha,\gamma)$ rate.  
   The abundance ratio between \nuc{20}Ne and \nuc{22}Ne results essentially from the competition between the $(\alpha, n)$ and $(n, \alpha)$ reactions of \nuc{17}O, and hence, $Y_{20} / Y_{22} = (1 - f) \xi$.  
   Then \nuc{22}Ne overweighs \nuc{20}Ne except for small mixing of $\dmixm \lesssim 0.001$ since $\xi$ decreases for greater mixing, and attains at the abundance of 
\begin{equation}
Y_{22} \simeq \Delta Y_{\rm 13, mix} / 2, \hbox{ or } [\abra{Ne}{C}] \simeq 0.21 + \log \dmixm
\label{eq:pro-ne}
\end{equation} 
unless the peak temperature reaches high enough to promote $\alpha$-captures of \nuc{22}Ne.
   Here we adopt the solar abundance of \citet{Asplund2005}.  

Since \nuc{23}Na is formed by neutron capture, the abundance is limited by the ratio of neutron cross sections, and from eq.~(\ref{eq:pro-ne}) we have 
\begin{equation}
Y_{23} \le 0.028 Y_{22} \simeq 0.014 \Delta Y_{13, \rm mix}, \hbox{ or } [\abra{Na}{C}] \le 0.33 + \log \dmixm, 
\label{eq:pro-Na}
\end{equation}
which may be compared with the enhancement during the third dredge-up in eq.~(\ref{eq:pro-Natdu}) of Appendix~2.  

\subsubsection{Magnesium isotopes and Aluminum}

As for the creation of \nuc{24}{Mg}, the difference between the $\nucm{20}{Ne}(\alpha, \gamma)$ and $\nucm{21}{Ne}(\alpha, n)$ reactions is not so large, giving $\lambda ^*_{\nucm{24}{Mg}}\simeq 30-700$ in Fig~\ref{fig:synthesis}.  
Furthermore, these reactions are slower than the 3$\alpha$ reaction for $\log T \lesssim 8.45$ (see Figure~\ref{fig:lifetimes}). 
   For weak flashes with such low peak temperatures as in Model~1, therefore, the dominant process to create \nuc{24}{Mg} is the neutron capture of \nuc{23}{Na} and the subsequent $\beta$-decay.  
   The magnesium isotopes keep increasing with Na, and eventually, \nuc{26}Mg becomes more abundant than \nuc{23}Na because of much smaller neutron capture cross section.  
   Under the steady-state condition, the ratio of magnesium to sodium may reach
\begin{equation} 
   Y_{26} / Y_{23} <  \sigma_{n \gamma} (\nucm{23}{Na}) / \sigma_{n \gamma} (\nucm{26}{Mg}) \simeq 16.7, \hbox{ or } [\abra{Mg}{Na}] < -0.12. 
\label{eq:pro-mg}
\end{equation}
Here $[\abra{Mg}{Na}]$ includes the contributions of all stable Mg isotopes.
For strong shell flashes with such high peak temperatures as in Model~5, the $\alpha$-captures may compete or overweigh the neutron captures, in particular when the mixing is weak and the neutron density is small.  
   Because of larger $\alpha$-capture rates of \nuc{22}Ne than that of the other neon isotopes, \nuc{25}Mg and \nuc{26}Mg are produced more than \nuc{24}Mg.  
   In this case, the \nuc{26}Mg has also the largest abundance among the magnesium isotopes, since \nuc{25}Mg captures neutron with much larger cross section. 
   But note that $\alpha$-capture reactions of \nuc{22}{Ne} are subject to large uncertainties about three orders of magnitude in the largest, as seen from the NACRE compilation. 
   Recently, the uncertainty has been reduced by the reanalysis but still remains as large as a factor of ten \citep{Koehler2002}.  

Consequently, the peculiar abundance ratios among the magnesium isotopes result from the difference in neutron capture cross sections as well as from the contribution of $\alpha$-capture reactions of \nuc{22}{Ne}.  
   The most abundant isotope is \nuc{26}Mg, while the other isotopes, \nuc{24}Mg and \nuc{25}{Mg}, turn out to be as abundant as \nuc{23}{Na}, reflecting the nearly similar neutron capture cross sections. 
   Because of the very small cross section of \nuc{26}Mg, an additional steep decrease in the abundances results between \nuc{26}Mg and \nuc{27}Al.  
   The upper limit of the ratio is given by 
\begin{equation}
Y_{27} / Y_{26} <  \sigma_{n \gamma} (\nucm{26}{Mg}) / \sigma_{n \gamma} (\nucm{27}{Al}) \simeq 1/ 30, \hbox{ or } [\abra{Al}{Mg}] < -0.34, 
\label{eq:pro-al}
\end{equation} 
where the contributions from all Mg isotopes are included in the [$\abra{Al}{Mg}$] ratio. 

\subsubsection{Si isotopes and heavier elements}

Beyond magnesium isotopes, the elements are created solely through a series of neutron captures and $\beta$-decays since the $\alpha$-capture reactions can no longer make any significant contribution under the temperature realized during helium shell flashes, as seen from Figure~\ref{fig:lifetimes}.  
   The fact expresses itself in the abundance distributions of Al through S isotopes in Figures.~\ref{fig:yields} and \ref{fig:mixtime-dep}. 
   All the models show a general trend that their abundances decrease for heavier nuclide and also for larger neutron capture cross sections. 
   The abundances increase with the amount of mixed \nuc{13}C more rapidly than linearly and approach the steady-state ones inversely proportional to the cross sections.  
  Accordingly, the silicon isotopes are less abundant than, or comparable to, Al, and local peaks appear at phosphorus and \nuc{34}S, which have much smaller neutron capture cross sections as compared with ambient nuclides.  

\section{Comparisons with the Observations of HMP/UMP and EMP Stars}

We discuss the relevance to the abundance patterns observed for the stars in the Galactic halo from the viewpoint of the binary scenario \citep{Suda2004}.  
   In this scenario, the nuclear products in the helium zone are eventually carried out to the surface of erstwhile primary stars as a result of the deep mixing and dredge-up (He-FDDM), which occurs after the shell flash grows strong enough to engulf hydrogen at sufficiently high rate. 
   In addition, the nuclear products are also dredged up to the surface during the helium shell-flashes (TDU) subsequent to He-FDDM in some cases.  
   Then, the matter in the envelope of the primary stars is planted on the low-mass secondary members through wind accretion to be observed as the surface pollution. 
   The yields may suffer dilution with the envelope matter both in the convective envelope of primary stars when they are dredged up and in the surface convection of secondary stars when they are accreted.  
   Accordingly, it makes sense to compare the relative abundances among the elements. 

\begin{figure}
\FigureFile(120mm,120mm){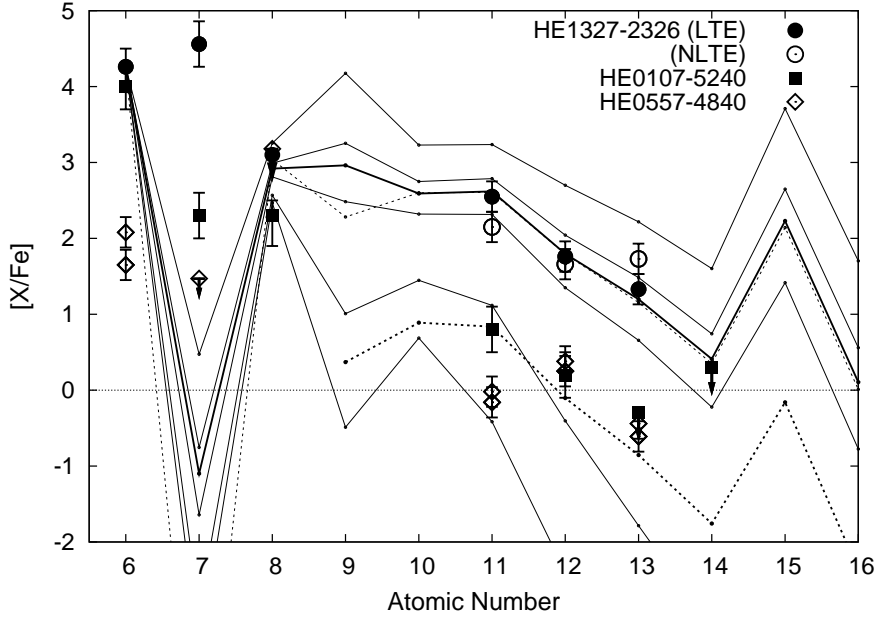}
\caption{
Comparison of the elemental abundances of yields of AGB nucleosynthesis resultant from hydrogen mixing into the helium flash convective zone with the observed abundance patterns of light elements for three most iron-deficient stars; 
the computational results are plotted, normalized with \nuc{12}C abundance of HE1327-2326 for Model~1 with the different amounts of mixed $^{13}$C, $\dmixm = 0.0001$, 0.001, 0.01, and 0.02, 0.03, and 0.1 and all with $\Delta t_{\rm mix} = 10^{12}$ s, by solid lines from bottom to top, respectively.
   The upper dotted line is the result for Model~3 of higher peak temperature with $\dmixm = 0.02$ and $\Delta t_{\rm mix} = 10^{10}$ s.
   The lower dotted line is obtained by multiplying the above result by $1/20$ under the assumption that the elements are diluted by a factor of 20 relative to the carbon abundance which increases due to the third dredge-up subsequent upon He-FDDM\citep[see][]{Suda2004}.
   The observed abundances and upper limits are taken from \citet{Christlieb2004} for HE 0107-5240, from \citet{Aoki2006} for HE1327-2326, and from \citet{Norris2007} for HE 0557-4840. 
   The oxygen abundance is taken from \citet{Bessell2004} for HE0107-5240, and the middle of three values (see Figure~\ref{fig:o-emp}) taken from  \citet{Frebel2006} for HE1327-2326, while only upper limit from \citet{Norris2007} for HE 0557-4840.
}
\label{fig:comparison2obs}
\end{figure}

Figure~\ref{fig:comparison2obs} shows the abundance patterns of yields of our computations for various amounts of mixing between $\dmixm= 10^{-4}$ and 0.1 (solid lines).  
   Note that the elemental abundances are plotted so as to give the same carbon enrichment relative to iron as observed for HE1327-2326 ($[\abra{C}{Fe}] = 4.26$).  
   In addition to carbon, oxygen is also enhanced as large as $[\abra{O}{C}]\simeq -2 \sim -1$ with rather weak dependence on the amount of mixed \nuc{13}C.  
   The neon enhancement is in proportion to the amount of mixed \nuc{13}C. 
   For heavier elements, the enhancement decreases in general and the dependence on the mixing amount grows stronger.  
   Sodium, magnesium, and aluminum are all enhanced for $\dmixm \gtrsim 0.01$ while only sodium for $\dmixm \simeq 0.001$.
   Fluorine and phosphorus are greatly enriched for $\dmixm \gtrsim 0.01$, reaching the comparable enhancement as carbon for $\dmixm \simeq 0.1$ because of their small solar abundances. 

In comparing the yields from helium shell flashes with the observed abundances, we should take into account some later modifications that may be introduced when and after they are dredged up to the surface.   
During He-FDDM, a hydrogen shell flash ignites in the middle of helium convective zone, and the largest effect is to burn carbon into nitrogen as much as $[\abra{N}{C}] \simeq 0.7 \sim - 0.3$ in the hydrogen-flash convection \citep{Fujimoto2000,Iwamoto2004}, while other heavier elements are mostly left unaltered because of larger Coulomb barriers and a short duration of hydrogen flash-convection. 
   Hydrogen mixing and He-FDDM works for stars of mass $M \lesssim 3 \msun$ until the CNO abundances in the envelope exceeds $[\abra{CNO}{Fe}] \gtrsim -2.5$ \citep{Fujimoto2000,Suda2008a}.  
   For stars of mass $M \gtrsim 1.5 \msun$, TDU follows He-FDDM to enrich the envelope with \nuc{12}C and \nuc{16}O but without nitrogen. 
   In the envelope of massive AGB stars, the nuclear products, after dredged up to the surface, are further processed by hydrogen burning in the envelope convection (hot bottom burning; HBB). 
   There are the lower limits of stellar mass for TDU and for HBB, and are argued to decrease for smaller metallicity. 
   It is true, however, that they are subject to uncertainties due to the treatment of convective mixing, such as the mixing length theory and overshooting.
   We rather consider that these parameters should be determined from the comparison with the observations of EMP stars. 

The observed abundances for three HMP/UMP stars are also plotted by symbols.  
    For HE1327-2326, the pattern of abundance enhancement from carbon through aluminum is well reproduced by our yields with the amount of mixing $\dmixm \simeq 0.02$ as long as the LTE abundances are concerned (filled circles). 
	Nitrogen enrichment ($[\abra{N}{C}]=0.3$; Aoki et al.~2006)is also consistent with the processing by hydrogen shell-flash during He-FDDM, as shown by \citet{Fujimoto2000} for $0.8 \msun$ with $Z = 0$ and $\feoh =-4$, and by \citet{Iwamoto2004} for $1 - 2 \msun$ and $\feoh = -2.7$, if we take into account the differences in the metallicity and the core mass. 
   The non-LTE correction (open circles) gives smaller abundance to sodium while larger abundance to aluminum than the LTE abundances both by $\sim 0.4$ dex, but remains within the 2$\sigma$ of typical errors.    
   
For HE0107-5240, the operation of TDU subsequent upon He-FDDM is proposed by \citet{Suda2004} to explain the small nitrogen abundance by a factor of $\sim 30$.  
   As the surface carbon abundance increases $\sim 30$ times by TDU, the surface nitrogen abundance is reduced as much but the surface abundances of the other yields are reduced by a factor of $\sim 20$ (dashed line).  
   The difference in the reduction factors is related to the fact that without hydrogen mixing, nitrogen is burnt out but the other yields can survive in the convective zone during the subsequent helium shell-flashes to be dredged up at the same time with carbon \citep[see][]{Suda2004}.  
   The effect of dilution by TDU is consistent with the observed sodium enhancement.  
   The magnesium abundance with rather small enhancement and an upper limit of aluminum abundance are also consistent with the dilution.  

As for HE0557-4840, the enhancement is observed only for carbon ($[\abra{C}{Fe}] = 1.7- 2.1$), but the excess is much smaller as compared with two HMP stars; 
   nitrogen and oxygen are given only the upper limits. 
   There are no indications for the enrichment for sodium, magnesium or aluminum \citep{Norris2007}. 
   The observed abundance pattern does not necessarily demand the hydrogen mixing events and is explicable in terms of TDU alone. 
 
In the following subsections, we discuss the comparisons on the carbon and oxygen ratios and the abundances of light elements of sodium, magnesium and aluminum in some more detail for the HMP/UMP stars, and also for EMP stars. 
   In our computations, we assume the metal-free stars, and yet, the results may be little affected by the presence of pristine metals as long as their contributions to neutron captures are much smaller than that of \nuc{16}O.  
   Since the pristine CNO elements are converted into \nuc{22}Ne in the helium convective zone, the latter condition is reduce to
\[ Y_{56} \ll Y_{16, 0} \sigma_{n \gamma} (\nucm{16}{O}) / \sigma_{n \gamma} (\nucm{56}{Fe}) \simeq 0.027 Y_{56, \odot} (Y_{16, 0}/ 0.01Y_{12})( X_{12, \heconv}/ 0.2).   \]
   Accordingly, the present results may be applicable to EMP stars of $\feoh < -2.5$, and we take them into account. 
   On the basis, we finally discuss the origin of three HMP/UMP stars.  .

\subsection{carbon-oxygen enrichment}  

Our results suggest that the helium shell flash with hydrogen mixing is an effective site for oxygen production.   
   The neutron-recycling reactions can bring the oxygen and carbon ratio up to eq.~(\ref{eq:ox-pro2}), and hence, 
\begin{equation} 
[\abra{O}{C}] \le -0.66,   
\label{eq:ocr-fddm}
\end{equation}
   with the use of the solar abundances by \citet{Asplund2005}, although it demands larger neutron exposure, or larger mixing of \nuc{13}C, than in our computation to realize the ratio close to it. 
   This upper bound is larger by a factor of $\sim 10$ than the ratio in eq.~(\ref{eq:ox-proag}) that can be attained by TDU alone; 
\begin{equation} 
[\abra{O}{C}] = (-2.5 \mathchar`- -1.75) + \log[(X_{12, \rm He\mathchar`-conv}/0.2) (0.8 / Y_{\heconv})^2 /(\rho / 3 \times 10^{3} \hbox{ g cm}^{-3})], 
\label{eq:ocr-tdu}
\end{equation}
   where $X_{12, \heconv}$ and $Y_{\heconv}$ denote the abundances of carbon and helium in the helium flash convective zone, respectively, and the density is taken from the middle stages of the strongest shell flashes (Models 4 and 5) since TDU occurs for relatively massive stars.    

\begin{figure}
\FigureFile(120mm,120mm){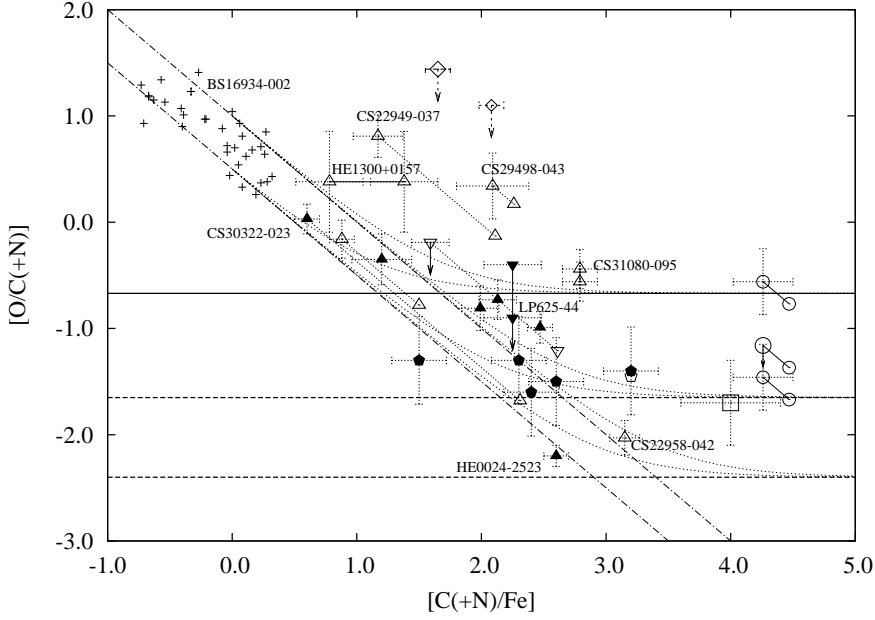}
\caption{
Comparison of the oxygen abundances, theoretically predicted from the nucleosynthesis in AGB stars, with the observations for HMP/UMP and EMP stars, where the ratios of the oxygen to carbon abundances are shown against the carbon enhancements; 
   thick solid line denotes the upper limit of the $[\abra{O}{C}]$ values, reached by hydrogen mixing and He-FDDM.
   Two broken lines bound the range that can be reproduced by TDU without hydrogen mixing.  
   Symbols denote the observed oxygen abundances for stars of $\feoh < -2.3$, taken from Stellar Abundance for Galactic Archeology Database \citep[SAGA Database;][]{Suda2008b};
   open circles and square denote HE1327-2326 and HE0107-5240, respectively and diamonds show the upper limits of oxygen abundances derived under two different effective temperatures for HE0557-4840 \citep{Norris2007}:  
   filled triangles and open triangles denote CEMP-$s$ and CEMP-no$s$, respectively, while filled and open pentagons show those based on the medium-resolution spectroscopy by \citet{Beers2007}: and crosses denote EMP stars without carbon enhancement ($[\abra{C}{Fe}] < 0.5$).  
   For HE1327-2326, plotted are three oxygen abundances, an upper bound derived for OI triplet line $\lambda 777$ nm (middle), an abundance by 1D LTE analysis of UV OH features (the highest one), and one with 3D correction (the lowest one) \citep{Frebel2006}.
   We show two data by 1D LTE analysis of UV spectra and with 3D corrections included for HE1300-0157 \citep[right and left ones, respectively][]{Frebel2007},  the upper limit derived from [OI] $\lambda 6300$ as well as the abundance from OI triplet for LP 625-44,\citep[lower upper ones, respectively][]{Aoki2002c}, and LTE oxygen abundances from OI triplet line $\lambda 777$ nm and that with NLTE correction for CS31080-095 \citep{Sivarani2006}.    
   For CEMP stars with the nitrogen enrichment larger than the carbon enrichment, the abundances where nitrogen is added to carbon are indicated by open triangles linked by dotted line.
   Two oblique thin solid lines denote the loci of $[\abra{O}{Fe} = 0.5]$ and 1.0, and thin broken lines delineate the ranges for a mixture of the yields of AGB nucleosynthesis and the envelope matter of the pristine oxygen abundance  $[\abra{O}{C}]=0.5$ and 1.0.
}
\label{fig:o-emp}
\end{figure}

Figure~\ref{fig:o-emp} shows the comparison of these theoretical predictions from the AGB nucleosynthesis and dredge-up with the observed ratios of oxygen to carbon for HMP/UMP and EMP stars. 
   The upper bound for He-FDDM in eq.~(\ref{eq:ocr-fddm}) and the upper and lower limits for TDU in eq.~(\ref{eq:ocr-tdu}) are drawn by thick solid and broken lines, respectively.    
   For HE1327-2326, we take the carbon abundance of $[\abra{C}{Fe}] = 4.26 \pm 0.24$ from the 1D LTE analysis of CH lines \citep{Aoki2006}.
   For the oxygen abundances, an upper bound of $[\abra{O}{Fe}] < 3.1$ is derived from non-detection of both OI triplet at $\lambda 7771-5$ and [OI] at $\lambda 6300$, while the abundances $[\abra{O}{Fe}] = 3.7 \pm 0.2$ and $2.8 \pm 0.2$ result from the 1D LTE analysis of UV OH features and from the applications of 3D corrections, respectively \citep{Frebel2006}. 
   These ratios are consistent with the range that can be reproduced by the hydrogen mixing and dredge-up by He-FDDM. 
   Even the largest one is consistent with our computation within an error if we regard nitrogen as carbon processed by hydrogen shell-flash during He-FDDM . 
   The ratios, including the upper limit for OI line, lies well above the range for TDU.  
   For HE0107-5240, we have the carbon and oxygen abundances only from the 1D LTE analyses at $[\abra{C}{Fe}] = 4.0 \pm 0.2$ \citep{Christlieb2004}, and $[\abra{O}{Fe}] = 2.3 \pm 0.2$ from UV OH lines \citep{Bessell2004}, which gives slightly smaller ratio of $[\abra{O}{C}] = - 1.7 \pm 0.3$ than for HE1327-2326.  
   This ratio lies just below the upper border of the range in eq.~(\ref{eq:ocr-tdu}), and is explicable in terms of TDU, if allowance is made for errors in the abundance determination.  

   For EMP stars, we plot all the samples that we find the measured oxygen abundances in the literature with use of SAGA Database \citep[The Stellar Abundance for Galactic Archeology Database;][]{Suda2008b}.   
   There is a general trend that the $\abra{O}{C}$ ratios decrease with carbon enrichment.  
   The largest ratios of $[\abra{O}{C}] \simeq 0.5 - 1.5$ are found for EMP stars without the carbon enhancement ($[\abra{C}{Fe}] < 0.5$), among which the oxygen abundances are in the range between $[\abra{O}{Fe}] \sim 0.5$ and 1.0 and the variations in the $[\abra{O}{C}]$ ratios are attributable to the conversion of carbon into nitrogen possibly by CN cycle processing \citep{Spite2005}.   

A salient feature of the oxygen abundances in the EMP stars is that such large enrichments as exceeding $[\abra{O}{Fe}] \simeq 1$ are observed only for the carbon-enhanced EMP (CEMP) stars of $[\abra{C}{Fe}] \ge 0.5$. 
Among the EMP giants without carbon enhancement, the largest abundance is $[\abra{O}{Fe}] = 1.14$ with the largest ratio $[\abra{O}{C}] = 1.41$ for BPS BS 16934-002 in this figure \citep[][see also SAGA Database]{Aoki2007b}.  
   The CEMP stars exhibit lower $\abra{O}{C}$ ratios, and most of them have such low ratios as explicable in terms of the AGB nucleosynthesis with hydrogen mixing and He-FDDM and/or with TDU. 
   In this figure, we delineate the ranges of $[\abra{O}{C}]$ for the various mixtures of the matter, dredged up by He-FDDM or by TDU with the envelop matter of the pristine oxygen abundance taken to be $[\abra{O}{Fe}] = 0.5$ or 1.0, as a function of carbon enrichment.   
   CEMP stars divide into two subgroups according to the s-process enrichment, i.e., CEMP-$s$ with $[\abra{Ba}{Fe}] \ge 1.0$ (filled symbols) and CEMP-no$s$ with $[\abra{Ba}{Fe}] \le 1.0$ (open symbols).
   All the CEMP-$s$ stars fall in the domain of $\abra{O}{C}$ ratios realized by He-FDDM and TDU, mostly in the range that cant be attained only by He-FDDM, and about a half of CEMP-no$s$ stars also exhibit small $\abra{O}{C}$ abundances consistent with the AGB nucleosynthesis.  
   Two of the smallest $\abra{O}{C}$ ratios (HE0024-2523, CS22958-042) are quite low values reproducible by TDU alone. 
   HE0024-2523 belongs to a binary of period 3.14 days, indicative that it has undergone the Roche-lobe overflow and the common envelope phase \citep{Lucatello2003}, while CS22958-042 is a CEMP-no$s$ star but shows the largest sodium enrichment \citep{Sivarani2006}.     
   In addition, for CS30322-023, which has a large nitrogen enrichment relative to carbon ($[\abra{N}{C}] = 2.21$; Masseron et al.~2006), the $\abra{O}{C}$ ratio reduces as small as consistent with the TDU nucleosynthesis without hydrogen mixing if nitrogen is regarded as having been produced from carbon ($[\abra{O}{C(+N})] \simeq -1.7$).   
   If this is the case, the observed large ratio of nitrogen to carbon demands a deep processing by CN cycle, and the most likely site is the hot bottom burning in the envelope of massive AGB stars.  
   This in turn supports the binary scenario along with the variations in the radial velocity of period 191.7 days, at variance with the argument by \citet{Masseron2006} that this star is a currently TP-AGB stars of mass $\sim 0.8 \msun$.  

   On the other hand, some CEMP-no$s$ stars show the $\abra{O}{C}$ ratios and the enrichment of $\abra{O}{Fe}$ too large to be explicable in terms of the AGB nucleosynthesis, and seem to comprise the groups that demand other oxygen source(s).  
   Among them are two giants, CS29498-043 and CS22949-037, both greatly enriched with nitrogen as $[\abra{N}{Fe}]=2.27$ \citep[CS29498-043;][]{Aoki2004} and 2.59 \citep[CS22949-037;][]{Depagne2002} as well as with Na and Mg greatly but not with Al (see below).  
   If nitrogen is added to carbon, the $\abra{O}{C}$ ratio prior to the processing by CN cycles may decrease and yet, are still too large to be reproduced by the AGB nucleosynthesis.  
   Another CEMP-no$s$ subgiant, HE1300+0157, exhibits a much larger oxygen enrichment, $[\abra{O}{C}]=0.4$ ($[\abra{O}{Fe}]=1.8$) than expected from the AGB nucleosynthesis, although the overabundances may be alleviated if 3D effect is included in the abundance analysis \citep{Frebel2007}.   
   This source is different from the above two in the absence of the enrichment of Na-Al.
   In addition, a CEMP-no$s$ dwarf, CS 31080-095, also exhibits a slightly larger enrichment of oxygen relative to carbon than expected from our results.  
   It may be distinct from the other three sources in larger metallicity by $\sim 1$ dex and also in a moderate enrichment of neutron capture elements ($[\abra{Ba}{Fe}] = 0.77$).  

In summary, the AGB nucleosynthesis with the hydrogen mixing in combination of pristine oxygen abundance of $[\abra{O}{Fe}] \simeq 0-1$ can provide reasonable explanation to the observed carbon and oxygen abundances for the majority of CEMP stars (all the CEMP-$s$ stars and half of CEMP-no$s$ stars) including HMP stars.  
In addition, our results may serve as a filter to single out such stars as demand other sources of carbon and oxygen, and identify the nucleosynthetic signatures presumably of Pop.~III and early-generation supernovae. 

\subsection{Na, Mg and Al enrichment}  

\begin{figure}
\FigureFile(120mm,120mm){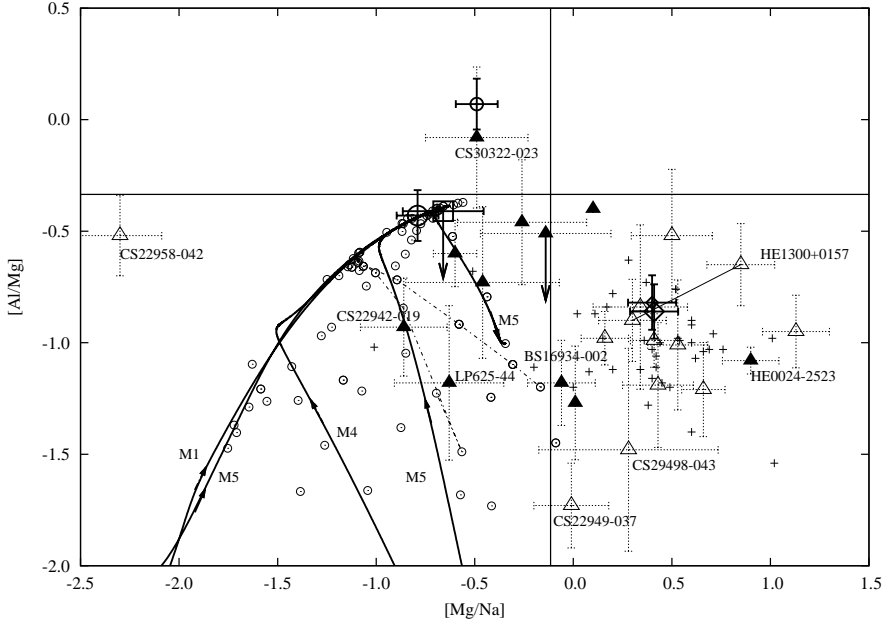}
\caption{
Abundance ratios among sodium, magnesium and aluminum, predicted from the nucleosynthesis of neutron capture reactions in the helium flash convection attendant upon hydrogen mixing and the comparisons with the observations for HMP/UMP and EMP stars on the $[\abra{Mg}{Na}]$ and$[\abra{Al}{Mg}]$ diagram. 
   Thick solid lines with arrow depict the evolutionary paths for some cases and small circles denote the end products of numerical computations.  
   Labels (M1 , M4, and M5) attached to solid lines indicate the results for Model 1 with $\dmixm =0.1$ and $\Delta t_{\rm mix} = 10^{12}$ s, for Model 4 with $\dmixm =0.1$ and $\Delta t_{\rm mix} = 10^{10}$ s and for two Models 5 with $\Delta t_{\rm mix} = 10^{6}$ s (left) and $\Delta t_{\rm mix} = 10^{9}$ s (middle with upward arrow) and both with $\dmixm =0.1$, respectively, and broken and dash-dotted lines connect the sequences of different $\Delta t_{\rm mix}$ with fixed $\dmixm =0.01$ for Models 3 (lower) and 5 (upper), respectively.  
   The observed abundances are collected with use of SAGA Database and denoted by the same symbols as in Figure~\ref{fig:o-emp}.
   Two horizontal and vertical dotted lines indicate the upper limits to the abundance ratios, $[\abra{Mg}{Na}]$ and $[\abra{Al}{Mg}]$, that can be reached solely by the neutron capture reactions. 
}
\label{fig:namgal}
\end{figure}

Among the light elements that can be enriched by the neutron-recycling reactions, a number of observations exist only for sodium, magnesium and aluminum. 
   Figure~\ref{fig:namgal} shows a summary of the abundances of these elements, obtained by our numerical computations, and the comparison with the observations of HMP/UMP and EMP stars in the $[\abra{Mg}{Na}]$ and $[\abra{Al} {Mg}]$ diagram.
   The numerical results (open circles) fall in the left-bottom quarter demarcated by vertical and horizontal lines, which denote the steady-state ratios under the neutron captures, respectively ($[\abra{Mg}{Na}] = -0.12$ and $[\abra{Al} {Mg}] = - 0.34$).  

In this figure, we also delineate the evolutionary paths of nucleosynthesis for several models by solid curves.  
Those curves which run most leftwards represent the evolutionary paths for which neutron captures dominate over $\alpha$-captures in producing magnesium isotopes, as seen for Model 1 with $\dmixm=0.1$ and $\Delta t_{\rm mix} = 10^{12}$ s and for Model 5 with $\dmixm=0.1$ and $\Delta t_{\rm mix} = 10^{6}$ s. 
   Along them, the ratios of $\abra{Mg}{Na}$ and $\abra{Al}{Mg}$ both increase and approach the steady-state values at the intersection of vertical and horizontal lines.  
   The approach to the steady-state abundance ratio is faster between Al and Mg than between Mg and Na since the ratio of the neutron capture cross sections of Al to Mg is much larger than that of Mg to Na, and hence, much smaller abundance ratio is sufficient for the former than for the latter to reach the steady state.
   The paths delineate the left boundary of the distribution of our yields which we call ``the neutron capture path'' in the following.  

   In comparison, some models run on the right side of the neutron-capture path, as seen for Model 5 with $\dmixm=0.1$ and $\Delta t_{\rm mix} = 10^{9}$ s and for Model 4 with $\dmixm = 0.1$ and $\Delta t_{\rm mix} = 10^{10}$ s.
   This is the case for the models with high peak temperatures and long mixing duration (or small neutron density).
   Higher temperature and smaller neutron density promote the $\nucm{17}{O} (\alpha, \gamma) \nucm{21}{Ne} (\alpha, n) \nucm{24}{Mg}$ reactions to make the contribution to the production of \nuc{24}Mg larger.
   Thus the models trace courses with larger $[\abra{Mg}{Na}]$ for given $[\abra{Al}{Mg}]$ in this figure.
   As \nuc{22}Ne accumulate and the neutron captures with \nuc{22}Ne as seed come to dominate over the Mg production, our models approach to the neutron-capture path and eventually join in it.

After the cease of mixing of \nuc{13}C, the $\alpha$-capture reactions proceed if the temperature remains sufficiently high.
For the models of short durations of mixing with the high peak temperatures, therefore, the evolutionary path departs again from the neutron-capture path and runs to right- and down-ward as magnesium isotopes are formed by $\alpha$-captures of \nuc{22}Ne. 
   The shorter the duration of mixing, the greater is the departure of the end products from the neutron-capture path. 
   Two dash-dotted lines denote the resultant abundance ratios for Model 5 (upper line) and Model 3 (lower line) both with $\dmixm = 0.01$.  
  
The HMP stars exhibit the abundances of Na, Mg and Al well within the range that can be reproduced by the neutron-capture reactions, which are quite different from the averaged values that are observed for EMP stars without the carbon enrichment ($[\abra{Mg}{Na}]\simeq +0.5$ and $[\abra{Al}{Mg}]\simeq -1$). 
  HE1327-2326 is located right on the neutron capture path when we adopt LTE abundance ratios.
  HE0107-5240 falls near but right to HE1327-2326 despite that the enrichments of Na and Mg are smaller by $\sim 1.6$ dex and more.  
   Since the Al abundance is an upper limit, the abundance ratio lies below the neutron-capture path, which is taken to be indicative of the shell flash of higher temperatures.  
   On the other hand, the UMP star, HE0557-4840, displays the absence of significant enrichments for any of these elements and is located in the middle of the EMP stars without carbon enrichment.  

The distribution of EMP stars on this diagram is clearly divided between CEMP-$s$ and CEMP-no$s$ stars.   
   All the CEMP-$s$ stars fall within, or near to, the ranges of $\abra{Mg}{Na}$ and $\abra{Al}{Mg}$ ratios, encompassed by the numerical results, except for the two stars of short-period binaries, HE0024-2523 \citep[period of 3.14days][]{Lucatello2003} and CS30322-023 \citep[possibly 192days][]{Masseron2006}, which are thought to have undergone the common-envelope phase. 
   This is consistent with the fact that the s-process nucleosynthesis is a necessary consequence of neutron capture reactions under the existence of pristine seed nuclei.    
   The observed values of $[\abra{Mg}{Na}]$ are shifted slightly rightwards of larger magnesium abundance from the neutron capture path.  
   This may be attributable to the contribution of pristine \nuc{24}Mg which has presumably larger abundance than sodium. 
   In addition, most of CEMP-$s$ stars have the sodium and carbon ratios similar to, or below, the value $[\abra{Na}{C}] \simeq -1.7$ observed for HE1327-2326.  
   Therefore, the origin of these elements is explicable in terms of hydrogen mixing of $\dmixm \lesssim 0.02$, while two stars, LP625-44 and CS22942-019, have larger sodium to carbon ratios ($[\abra{Na}{C}]=-0.50$ and $-0.78$). 

In contrast, CEMP-no$s$ stars distribute on the bottom-right quarter in this figure with smaller Na abundance relative to Mg than predicted from the neutron-capture nucleosynthesis; 
   most have the abundance ratios of $[\abra{Al}{Mg}]$ and $[\abra{Mg}{Na}]$ similar to EMP stars without the carbon enhancement, but with larger scatters. 
   The two CEMP-no$s$ stars with large $\abra{O}{C}$ ratios, CS29498-043 and CS22949-037, discussed above, are characterized by the large enhancement of Na and Mg. 
   \citet{Depagne2002} and \citet{Aoki2004} discuss the peculiar abundance patterns of these stars in connection with a peculiar supernovae such as proposed by \citet{Umeda2003} and also with the rotational effects as asserted by \citet{Meynet2002}.
   These stars have the $\abra{Al}{Na}$ ratios similar to the CEMP-$s$ stars but the magnesium abundances are larger by $\sim 1$ dex.
   The comparison with the evolutionary paths of Model 5 in the figure suggests the nucleosynthesis during helium shell flashes of larger core masses with smaller neutron density for which the temperatures reach high enough to promote the $\alpha$-capture of \nuc{22}Ne. 
   \citet{Komiya2007} propose that CEMP-no$s$ stars are formed from the binaries with a primary star of mass $M > 3 \msun$ and also with still more massive primary that develops an ONe core and ends with a supernova explosion.  
   The binary origin with the massive primary can be an alternative, while we have to seek for the source of large oxygen enrichment relative to the carbon enrichment.     
   The abundance patterns of these stars wait for further study, both theoretical and observational, which may shed light on the new aspect of the star formation and chemical evolution in the dawn of the universe.   

It is worth noting that the stars, CS22958-042, CS30322-023, and HE0024-2523, with such low $\abra{O}{C(+N)}$ ratios as consistent with TDU alone, display quite different behaviors in this diagram from each other; 
the largest enrichment of sodium ever known for CS22958-042 \citep{Sivarani2006}, large enrichment both of aluminum and sodium for CS30322-023 and no enhancement either of sodium nor aluminum for HE0024-2523. 
   While they share the nitrogen enrichment ($[\abra{N}{Fe}] = 2.15$, 2.81, and 2.1), the nitrogen to carbon ratios differs greatly ($[\abra{N}{C}] = -1.0$, 2.21 and $-0.5$).  
   The enhancement of s-process elements is absent for CS22958-042, while very large for HE0024-2523 and CS30322-023 with large $[\abra{Pb}{Ba}](=1.84, 0.95)$, respectively.  
   These may reflect the range of AGB nucleosynthesis according to the common-envelope evolution since these all are suspected as the members of short period binary systems from the radial velocity variations.

Although we have discussed so far about the LTE abundances, it is pointed out that departures from LTE have a significant impact on the abundance determination by use of the resonant lines. 
For HE1327-2326, the NLTE corrections give the $\abra{Al}{Mg}$ ratio larger than that of LTE abundances by $\sim 0.4$ dex and exceeding the upper limit reached via neutron capture reactions. 
   In order to explain the NLTE ratio within the present framework of neutron capture reactions, we need as much smaller ratio of neutron capture cross sections, $\sigma_{n \gamma} (\nucm{27}{Al}) / \sigma_{n \gamma} (\nucm{26}{Mg})$, which possibility is not excluded from the current uncertainties of their cross sections.  
   Otherwise, the enrichment of Al may be realized by proton captures with the burning of only a small fraction of Mg, but we have to seek for the appropriate site.   
   It is thought, however, that the abundance determination is subject to systematic errors also due to the 3D effects and this may have opposite effects to NLTE corrections \citep{Collet2006}.  
   We had better wait for future studies to consider the effects of departures from LTE and the uncertainties of nuclear reaction rates.

\subsection{Origin of Abundance Patterns of UMP/HMP Stars} 

We may now discuss the constraints on the origin of HMP/UMP stars within the framework of the binary scenario \citep{Suda2004}, based both on the above discussion on the abundance patterns and on the general picture of evolution of metal-free and extremely metal-poor stars, presented by \citet[][see also Komiya et al.~2007 and Suda \& Fujimoto~2008]{Fujimoto2000}.  

For HE1327-2326, the observed abundance pattern of the light elements from carbon through aluminum can be well reproduced by the neutron-recycling reactions in the metal-free or extremely metal-poor AGB stars.  
   The abundance ratios among Na, Mg and Al reflect the yields of neutron-capture reactions for the model of the lowest temperature with $\dmixm \simeq 0.02$, and nitrogen is produced during He-FDDM.
   The low-mass nature of the primary star is in accordance with the absence of the dilution due to TDU for these elements, which suggests that HE1327-2326 was formed as, and evolved from, the second member of a low-mass binary system with a primary star of mass $1.0 \msun \lesssim M \lesssim 1.5 \msun$.  
   For such a low-mass primary, it is possible to enrich the envelope with carbon up to, or even beyond, the observed enrichment of $[\abra{C}{H}] = -1.4$ by He-FDDM without TDU \citep{Fujimoto2000,Iwamoto2004}.   
   In addition, our interpretation is consistent with the observed depletion of Li since Li is readily destroyed during the AGB phase of primary star even if it has been formed during a hydrogen shell-flash of He-FDDM; 
   The s-process nucleosynthesis and Sr enrichment will be discussed in a subsequent paper.    

For HE0107-5240 abundances \citep{Christlieb2004}, the large enrichment of carbon ($[\abra{C}{H}] =-1.28$) and the sodium enrichment ($[\abra{Na}{Fe}] = 0.81$) along with the small nitrogen enrichment $[\abra{N}{C}] = -1.72$ to $-1.13$ may require both He-FDDM and TDU \citep{Suda2004}. 
   Accordingly, HE0107-5240 evolved from a binary system with the primary of mass between $1.5 \msun \lesssim M \lesssim 3 \msun$.  
   The oxygen abundance is explicable either by the hydrogen mixing or by TDU.  
   The sodium enrichment itself is relatively small in relation to carbon ($[\abra{Na}{C}]= -3.19$) and explicable by TDU alone as seen from eq.~(\ref{eq:pro-Na}).  
   With such a small pristine abundance ($\feoh = -5.3$ with normal CNO abundance), however, TDU cannot grow strong enough to enhance the carbon abundance up to the observed value \citep{Suda2008a}.  
   Accordingly, it must have experienced the hydrogen mixing and resultant He-FDDM to increase the carbon abundance above $[\abra{C}{H}] \gtrsim -2.5$. 
   \citet{Suda2004} predict the binary period of this system at $P_{\rm orb} = 76 \mathchar`-150$ yrs, and for HE1327-2326, the binary period can be even longer because of shallower surface convective zone \citep{Komiya2007}.

As for HE0557-4840, few information is available except for the relatively small carbon-enrichment of $[\abra{C}{H}] = -3.1 \mathchar`- -2.63$, smaller by an order of magnitude than the other two stars. 
The upper limit of nitrogen abundance excludes the low-mass primary star with He-FDDM since the ratio of N/C has to be significantly smaller than observed from HE1327-2326. 
   Differently from HE0107-5240, the small carbon enrichment is more likely to be attributed to the weak TDU in the primary star at such low metallicity \citep[see][]{Suda2008a}.  
   Thus we interpret that this star has evolved from a binary system with a primary of mass $M > \sim 3 \msun,  $ where the upper mass limit is set by the condition that dredged-up carbon can survive the hot bottom burning that are expected to work for massive AGB stars. 
   It is to be noted that the stars of mass $M > 2.5 \msun$, even if born of primordial gas, start the TP-AGB evolution in the helium shell polluted by metals accreted on the surface since the surface pollutants have spread all over the hydrogen-rich envelope at the second dredge-up when the surface convection penetrating into the helium core \citep{Suda2004}.  

In conclusion, the peculiar abundance patterns of light elements Li through Al, observed from the three HMP/UMP stars can be explained in terms of the nucleosynthesis in AGB stars of population III.   
The three stars represent all three possible ways for the metal-free and EMP stars of low- and intermediate-mass to acquire the carbon-enrichment through helium shell-flashes, formulated by \citet{Fujimoto2000}; 
   Case II with He-FDDM only, Case II$^\prime$ with He-FDDM plus TDU and Case IV of TDU only.  
   This gives a basis of the first-star interpretation of these stars, in which the enhancement of carbon and light elements is transferred from the erstwhile AGB companions and the other metals including the iron group elements are accreted from the interstellar gas, polluted with the supernova ejecta after their birth.  
   It is true, however, that the pristine metals of such small level with the scaled-solar abundances relative to the observed iron abundances have little effects on the evolution and nucleosynthesis of primary stars and these stars can develop the observed enhancement of light elements through the binary mass transfer. 
  Accordingly, we cannot exclude the possibility that these stars are the second-generation of stars, born out of gas already polluted by ejecta from the supernova explosion of the first generation of stars. 
   One of the ways to discriminate these two possibilities of iron sources is the s-process nucleosynthesis as discussed in \citet{Suda2004}.  
   As seen from Figure~\ref{fig:n-exp}, the neutron exposure well exceeds $\tau_{\rm exp} \simeq 3-4 \hbox{ mb}^{-1} (\dmixm/0.01)^{1/2}$, and hence, the heavy s-process elements can be synthesized even when initially the pristine metals as seeds are initially absent for sufficiently large amount of mixing.  
   The pristine iron, if exists, also undergoes the s-process nucleosynthesis and is mostly converted into lead with such large exposure. 
   In actuality, the enhancements of Sr and the upper limit of Ba are observed for HE1327-2326.  
   We will discuss this problem in a subsequent paper (Nishimura et al.\ 2009 in preparation).  

\section{Conclusions and Discussion}

A distinctive feature of metal-free (Pop.~III) and extremely metal-poor (EMP) stars of $\feoh \lesssim -2.5$ is the occurrence of hydrogen mixing into the helium convective zone during helium shell flashes in an early phase of TP-AGB phase, which promotes the neutron-capture reactions with \nuc{13}C formed via the capture of mixed proton by \nuc{12}C as neutron source.
We have studied the nucleosynthesis of light elements up to sulfur (\nuc{35}S), attendant with this hydrogen mixing and revealed the characteristics of AGB nucleosynthesis for Pop.~III and EMP stars.
   We demonstrate that they are the effective sites for the synthesis not only of carbon but also of oxygen and other light elements including the neutron-rich isotopes and odd elements.
   The resultant yields are compared with the abundance patterns observed for the EMP stars with carbon enhancement in the Galactic halo.  
   The present results serve to identify the trace of AGB nucleosynthesis in Pop.~III and EMP stars.  
   It is also useful to single out the nucleosynthetic signatures of other mechanisms, in particular, of the first supernovae.

Our main conclusions are summarized as follows; \hfill\break
(1) Under the neutron-rich environment, the neutron and $\alpha$-captures cooperate effectively to synthesize a variety of elements, starting from \nuc{12}C.  
   The combination of $(n, \gamma)$ and $(\alpha, n)$ reactions produce \nuc{16}O, \nuc{20}Ne, and \nuc{24}Mg, in much more abundances than the $(\alpha, \gamma)$ reactions.  
   Neutrons, released via $\nucm{13}{C} (\alpha, n) \nucm{16}{O}$, are spent mostly for \nuc{22}Ne production, and some for \nuc{26}Mg in the case of the strong flashes of large peak temperatures. 
   The neutron capture reactions proceed with \nuc{22}Ne and \nuc{26}Mg as seeds to form heavier elements including odd nuclei of sodium, aluminum and phosphorus. 
   The amounts of yields are determined essentially by the amount of mixed \nuc{13}C, with stronger dependence for heavier elements.  
   The abundance of \nuc{16}O increases in proportion to a square root of the amount of mixed \nuc{13}C until it reaches a significant fraction of \nuc{12}C, while the abundance of \nuc{22}Ne is in proportion to the amount of mixed \nuc{13}C and attains at a half of mixed \nuc{13}C in number.
   There is a general tendency of yields decreasing more rapidly for heavier elements while local peaks form in the abundance distribution at the nuclei with the neutron capture cross section much smaller than those for the ambient nuclei as \nuc{22}Ne, \nuc{26}Mg and \nuc{31}P.  

   (2) For the three most iron-poor stars, their peculiar abundance patterns from carbon through aluminum can be reproduced in terms of three possible modes of AGB nucleosynthesis of Population~III stars born of primordial gas for the primary stars of different mass range: 
   in a star of mass range $M \lesssim 1.5 \msun$ with helium shell flashes with hydrogen mixing and dredge-up by helium-flash driven deep mixing (He-FDDM) for HMP star of HE1327-2326 ($\feoh = -5.4$):
   in a star of mass range $\sim 1.5 \lesssim M_1 \lesssim 3 \msun $ with the additional nuclear products of helium shell flashes without hydrogen mixing and third dredge-up (TDU) following He-FDDM for the other HMP star of HE0107-5240 ($\feoh = -5.3$): 
   and in a star of mass in the range $M > 3 \msun$ with TDU alone for UMP star of HE 0557-4840 ($\feoh =-4.75$).

(3) The oxygen enhancements, observed for carbon enhanced EMP (CEMP) stars, are well explained in terms of the nucleosynthesis in the helium shell flashes with hydrogen mixing and subsequent helium flash-driven deep mixing (He-FDDM) or even without hydrogen mixing and subsequent third dredge-up (TDU). 
   This is the case in particular for all the CEMP-$s$ stars with $s$-process enhancement $[\abra{Ba (+ Pb)}{Fe}] \ge 1$ (CEMP-$s$ stars), while some CEMP-no$s$ stars without $s$-process enhancement $[\abra{Ba}{Fe}] < 1$ (CEMP-no$s$) exhibit the oxygen enhancement exceeding those predicted from the AGB nucleosynthesis. 
   As for the sodium, magnesium and aluminum, the observed abundances of CEMP-$s$ stars are accommodated within the neutron-capture nucleosynthesis during helium shell flashes with hydrogen-mixing along with the contribution of pristine magnesium, while CEMP-no$s$ lack the symptoms of the profound neutron captures.  
   
The agreement of the abundance characteristics for three HMP/UMP stars give the basis of our Pop.~III binary interpretation of their origins, i.e., these stars emerge from the low-mass members of binary systems of the first generation stars with the different masses, suffering the surface pollution by accreting the wind matter from the primary AGB stars and also by accreting interstellar matter enriched with iron and other metals ejected by supernovae after their birth.
   The three stars represent the entire possible paths to carbon-enrichment in the binary scenario, as formulated in \citet[][see Suda et al.\ 2004, and Komiya et al.\ 2007]{Fujimoto2000}; 
  HE1327-2326 for case II, HE0107-5240 for case II$'$, and HE0557-4840 for case IV.
   Our interpretation is consistent with the depletion of lithium in these stars.   
   It is true that the pristine CNO elements, if they exist as much as the scaled-solar abundances with the observed iron abundances of $\feoh = -5.4 \mathchar`- -4.75$, have nothing to do with the AGB nucleosynthesis as long as the light element synthesis is concerned.  
  And yet it is worth pointing out that the degree of the surface iron pollution is in accordance with our assignment of the masses to their primary stars since the Bondi accretion rate increase in proportion to the second power of total mass.  

One of the merits of our scenario is that it is worked out within the current standard framework of stellar evolution in spherically symmetry; 
   the stars of assigned masses must have brought about the characteristic abundances observed from these stars as long as they were formed.  
   The prevalence of binarity is known for younger populations and we may take it for granted among the stars of smaller metallicity \citep{Komiya2007}.  
   This forms a contrast with the supernova scenario, which assumes very artificial mixing-and-fallback model for supernova explosion, proposed by \citet{Umeda2003}. 
   Recently, \citet{Tominaga2007,Tominaga2009} attempt to interpret the mixing-and-fallback model as alleged jet-induced explosion, although it largely reduces the amount of carbon ejected and available for the formation of the next generation stars. 
   For these models to work, there should be a very efficient mechanism to confine most of the carbon ejected by a supernova and form a star in a cloud of rather small masses \citep[see, e.g.,][]{Suda2004}.  
   We also point out that the evolution of rotating stars lacks the sound foundations since a reliable theory is yet to be properly established for the hydrodynamical instability and elemental mixing in rotating stars.  

In the present study, we treat the mixing process parameterized with the amount of mixed \nuc{13}C and the duration of mixing epoch.  
   As seen in eq.~(\ref{eq:mix-hyd}), the mixing of \nuc{13}C, and hence, of hydrogen, is a small amount and at low rate not to cause the splitting of convective zone and He-FDDM. 
   Such weak hydrogen mixing is likely to be missed in actual numerical computations since the entropy in the hydrogen burning shell starts as the helium flash convection keeps increasing after the peak of shell flashes \citep{Fujimoto1977}.  
   In addition, the detailed modeling of hydrogen mixing will demand the proper treatment of the structure of convective motions and the entrainment of mixed hydrogen by the convection, which may be beyond the current framework of modeling of stellar structure and evolution.  
   Rather, we may gain an insight into the proper treatment from the comparison with the observations of EMP stars. 
   The same prescription is also applicable to the hot bottom burning. 
   It is argued that the hot bottom burning grows efficient for lower metallicity, and some evolutionary calculations predict that carbon and oxygen are deeply processed by CNO cycles \cite[e.g.,][]{Campbell2008}. 
   On the other hand, it is true that there are only a few stars observed with the large $\abra{N}{C}$ ratios \citep{Johnson2007}.  
   Since the temperature at the bottom of surface convection depends sensitively on the treatment of convection, and hence, the constraints may be imposed from the comparison with the observations of EMP stars themselves.  
   The effects of such internal mixing, whether they can be treated within the current standard framework of stellar evolution and beyond it, wait for future works, and have to be explored by recourse to comparisons with detailed abundances observed for individual stars.
   The present results may contribute to the theoretical effort in this line providing a basis for the understandings the characteristics of EMP stars.

\bigskip

T.~N. is supported by JSPS Research Fellowship for Young Scientists, and M.~A. acknowledges the financial support of the FNRS (Belgium).  
T.~S. is supported by a Marie Curie Incoming International Fellowship of the European Community FP7 Program (contract number PIIF-GA-2008-221145).
  This work is based on one of the author's (T.~N.) dissertation submitted to Hokkaido University.  
  It is supported in part by a Grant-in-Aid for Science Research from the Japanese Society for the Promotion of Science (grant 15204010,18104003,19740098), and in part by the Interuniversity Attraction Pole IAP 5/07 of the Belgian Federal Science Policy and by the Konan University - Universit\'e Libre de Bruxelles Convention `Construction of an Extended Nuclear Database for Astrophysics'.

\appendix

\section{Neutron-Recycling Reactions and Oxygen Production in Metal-Poor AGB Stars}

If we focus on the allocation of neutrons produced by the $(\alpha, n)$ reactions, the progress of neutron-recycling reactions may be described in the form 
\begin{equation}
   {d Y_{16} \over d \tau } = Y_{12} \sigma_{n \gamma} (\nucm{12}{C}) - Y_{16} \sigma_{n \gamma} (\nucm{16}{O}) 
\label{eq:O16A} 
\end{equation}
   as a function of the neutron exposure, $\tau$.  
   Here we may leave out \nuc{13}C since it takes a minor part as the destination of neutrons although the timescale is determined by the $\nucm{13}{C} (\alpha, n) \nucm{16}{O}$ reaction. 
   Since \nuc{12}C is much more abundant than other nuclides in our case, we may well neglect the variation in $Y_{12}$, and eq.~(\ref{eq:O16A}) yields an approximate solution analytically; 
\begin{equation}
Y_{16} = Y_{12,0} {\sigma_{n \gamma}(\nucm{12}{C}) \over \sigma_{n \gamma} (\nucm{16}{O})} \left[1 - \exp \{ - \sigma_{n \gamma} (\nucm{16}{O}) \tau \} \right] + Y_{16,0} \exp \{ - \sigma_{n \gamma}(\nucm{16}{O}) \tau\}. 
\label{eq:o16-sol}
\end{equation}  
   For $ \tau \ll 1/\sigma_{n \gamma} ( \nucm{16}{O} )$, then, the production of \nuc{16}O increases linearly with the neutron exposure as $\delta Y_{16} = Y_{16} - Y_{16,0} \exp [-\sigma_{n\gamma} (\nucm{16}{O}) \tau] \simeq Y_{12,0} \sigma_{n\gamma} (\nucm{12}{C}) \tau$, which is true mostly in our computations.  

   The flow of nucleosynthesis bifurcates on the passage through \nuc{17}{O} owing to the competition between the neutron and $\alpha$-captures, which may vary with the rate of hydrogen mixing and with the strength of shell flash.
   We denote the ratio between the neutron and $\alpha$-capture rates of \nuc{17}O by $\xi$ and the branching ratio of $(\alpha, \gamma)$ and $(\alpha, n)$ reactions of \nuc{17}O by $f$, i.e., 
\begin{eqnarray}
\xi & = & Y_{4} \{ \langle \sigma_{\alpha n} (\nucm{17}{O}) v \rangle +  \langle \sigma_{\alpha \gamma} (\nucm{17}{O}) v \rangle \} / Y_n \langle \sigma_{n \alpha} (\nucm{17}{O}) v \rangle,  
\label{eq:a-o-n-ratio}\\
f & = & \langle \sigma_{\alpha \gamma} (\nucm{17}{O}) v \rangle / \left( \langle \sigma_{\alpha n} (\nucm{17}{O}) v \rangle + \langle \sigma_{\alpha \gamma} (\nucm{17}{O}) v \rangle \right). 
\label{eq:branch-ratio}
\end{eqnarray} 
   Then, we may approximate the further progress of nucleosynthesis up to \nuc{22}Ne as
\begin{eqnarray}
{d Y_{17} \over d \tau} & = & Y_{16} \sigma_{n \gamma} (\nucm{16}{O}) - Y_{17} \sigma_{n \alpha} (\nucm{17}{O})  ( 1 + \xi), 
\\
{d Y_{14} \over d \tau} & = & Y_{17} \sigma_{n \alpha} (\nucm{17}{O}) , \\
{d Y_{20} \over d \tau} & = & Y_{17}  \sigma_{n \alpha} (\nucm{17}{O}) \xi (1 - f) - Y_{20} \sigma_{n \gamma} (\nucm{20}{Ne}) , 
\\
{d Y_{21} \over d \tau} & = & Y_{17} \sigma_{n \alpha} (\nucm{17}{O}) \xi f + Y_{20} \sigma_{n \gamma} (\nucm{20}{Ne}) - Y_{21} \sigma_{n \gamma} (\nucm{21}{Ne}) , 
\\
{d Y_{22} \over d \tau} & = & Y_{21} \sigma_{n \gamma} (\nucm{21}{Ne})
\end{eqnarray}
   where we neglect the absorption of neutron by \nuc{14}C because of smaller cross section. 
   For the weak shell flashes of longer timescales, \nuc{14}C may decay into \nuc{14}N that has much larger neutron capture cross section.  
   This makes little difference, however, since \nuc{14}N reacts mainly through $(n, p)$ channel and the released protons are converted again into neutrons through the $\nucm{12}{C} (p, \gamma) \nucm{13}{N} (e^+ \nu)\nucm{13}{C} (\alpha, n) \nucm{16}{O}$ reactions.  

\begin{figure}
\FigureFile(120mm,120mm){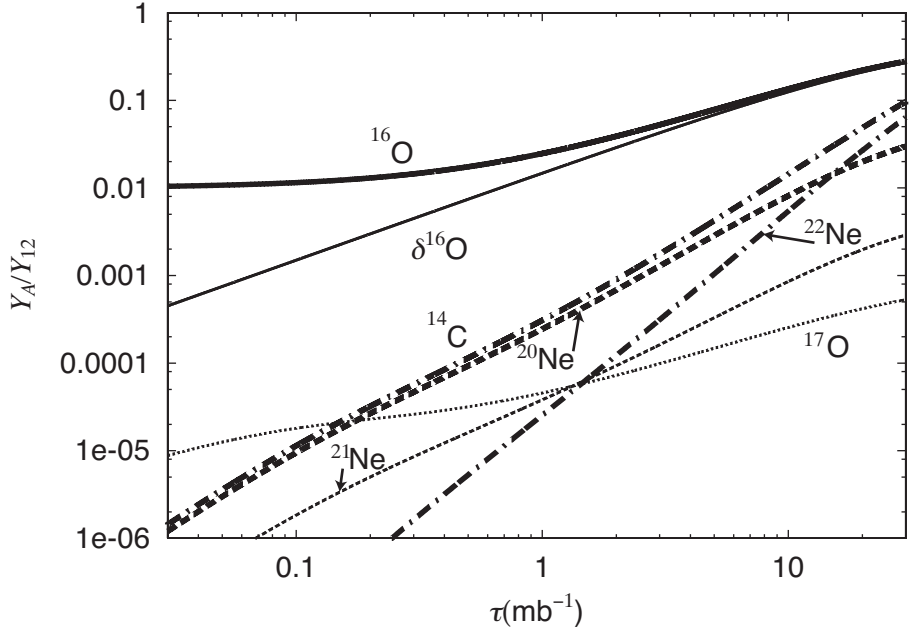}
\caption{
Progress of nucleosynthesis in the helium flash convection with the neutron capture reactions, plotted against the neutron exposure $\tau$.  
Variations of abundances are computed with use of the system of approximate equations given in Appendix 1 by assuming that the ratio between the neutron and $\alpha$-capture rates of \nuc{17}O equals to unity, i.e., $\xi = 1$ [see eq.~\ref{eq:a-o-n-ratio}];  
   thin solid labeled $\delta^{16}{\rm O}$ denotes the production of \nuc{16}O, $\delta Y_{16} = Y_{16} - Y_{16,0}\exp [-\sigma_{n\gamma} (\nucm{16}{O}) \tau]$.  
   Note that the $\alpha$-captures of \nuc{14}C is neglected, and if they are included, \nuc{14}C will be converted into \nuc{22}Ne. 
}
\label{fig:OCNe-tau}
\end{figure}

If we regard $\xi$ and $f$ as constant, neglecting the variations of the reaction rates with the temperature, the above system of equations yields an analytical solution.     
   Figure~\ref{fig:OCNe-tau} shows the evolution of number abundances of ${}^{16, 17}$O, \nuc{14}C and ${}^{20,21,22}$Ne for $\xi = 1$ as a function of neutron exposure; 
   this reproduces the characteristics observed from the numerical computations in Figure~\ref{fig:synthesis} very well.  
   The abundance of \nuc{16}O increases linearly with the neutron exposure, as stated above, and approach an upper bound of $Y_{16}/Y_{12} = \sigma_{n \gamma} (\nucm{12}{C}) / \sigma_{n \gamma} (\nucm{16}{O})$ for $\tau \gtrsim 1/\sigma_{n \gamma} ( \nucm{16}{O} )$. 
   In comparison, $Y_{17}$ and $Y_{14}$ remain in much smaller abundances.   
   They increase with $\tau$ in proportion to higher powers as 
\begin{eqnarray}
Y_{17} & \simeq &  \sigma_{n \gamma} (\nucm{16}{O}) [Y_{16,0}\tau + (1/2) Y_{12,0} \sigma_{n \gamma} (\nucm{12}{C}) \tau^2 ] 
\\
Y_{14} & \simeq & \sigma_{n \alpha} (\nucm{17}{O}) [(1/2) Y_{16,0} \sigma_{n \gamma} (\nucm{16}{O}) \tau^2 + (1/6) Y_{12,0} \sigma_{n \gamma} (\nucm{12}{C}) \sigma_{n \gamma} (\nucm{16}{O}) \tau^3] 
\end{eqnarray}    
   for small $\tau$.  
   For $\tau \gtrsim 1/ \sigma_{n \alpha} (\nucm{17}{O}) (1 + \xi)$, the former tends to reach a steady-state value, $Y_{17} \simeq Y_{12} \sigma_{n \gamma} (\nucm{12}{C}) / \sigma_{n \alpha} (\nucm{17}{O} ) (1 + \xi)$. 
   On the other hand, $Y_{14}$ keeps increasing simply because of the neglect of $\alpha$-capture of \nuc{14}C.   
   In actuality, \nuc{14}C as well as the daughter nuclei \nuc{14}N eventually burn into \nuc{18}O and into \nuc{22}Ne since the $\alpha$-captures of these nuclei and of \nuc{18}O are much faster than the 3$\alpha$ reactions (see Figure~\ref{fig:lifetimes}).    
   
The production of \nuc{20}Ne and \nuc{21}Ne is in proportion to $\xi$.  
Between them, \nuc{20}{Ne} is much more abundant than \nuc{21}{Ne} because of larger branching ratio $(1-f)/f$ and smaller neutron capture cross section of \nuc{20}Ne.  
   For large neutron exposure, their abundance ratio approaches a steady-state value of $Y_{20} (1-f) \sigma_{n \gamma} (\nucm{20}{Ne}) \simeq Y_{21} \sigma_{n \gamma} (\nucm{21}{Ne})$, and eventually, \nuc{22}Ne overweighs them.  
   Note that \nuc{22}Ne plotted in this figure is only those formed solely through neutron captures while much more is formed via the $\alpha$-captures of \nuc{14}C.

In the recycling reactions, neutrons are spent mostly to form neutron-rich isotopes of oxygen and neon so that the progress of nucleosynthesis is limited by the number of available neutrons, and hence, by the amount of mixed \nuc{13}C.  
   Accordingly, the termination is given by the relation
\begin{equation}
\Delta Y_{\rm 13, mix} = Y_{17} + Y_{21} + 2 \times (Y_{14} + Y_{22}). 
\end{equation}
   More neutron may be released from \nuc{17}{O} and neon isotopes.  
   The former is effective when the $\alpha$-capture overweighs the neutron capture and the latter are limited to the shell flashes of high temperatures ($T \gtrsim 3 \times 10^8$ K).  

\begin{figure}
\FigureFile(120mm,120mm){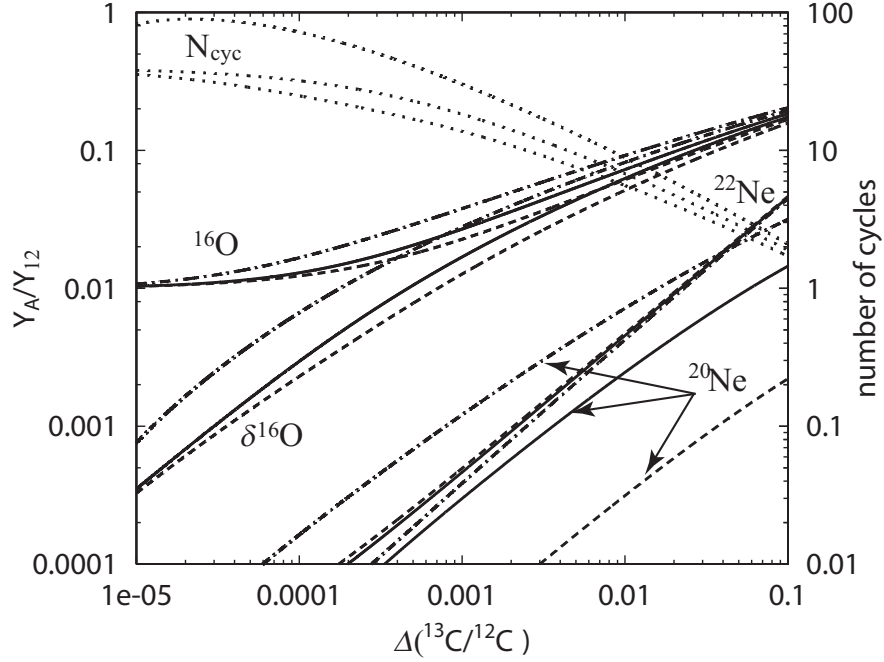}
\caption{
Production of oxygen and neon by the doubly neutron-recycling reactions as function of the mixed amount, $\dmixm$, of mixed \nuc{13}C relative to \nuc{12}C [$ = \Delta X_{13, \rm mix} / X_{12} (t_{\rm peak})$] for the different ratio, $\xi$, between the $\alpha$ and neutron capture reactions of \nuc{17}O.  
   Broken, solid, and dash-dotted lines denote the results for $\xi =0.1$, 1, and 10, respectively. 
   Dotted lines on the top denote the number, $N_{\rm cyc}$, of neutron recycling reactions $\nucm{12}{C} (n, \gamma) \nucm{13}{C} (\alpha, n) \nucm{16}{O}$ that each neutron experiences, and hence, the ratio of the produced oxygen to the mixed \nuc{13}C ($\delta \nucm{16}{O} / \Delta Y_{\rm 13, mix}$) for $\xi =0.1$, 1, 10, from bottom to top.  
}
\label{fig:Opro-dmix}
\end{figure}

   Figure~\ref{fig:Opro-dmix} shows the final abundances of \nuc{16}O and neon isotopes, \nuc{20}Ne and \nuc{22}Ne, where all \nuc{14}C are assumed to be converted into \nuc{22}Ne by the $\alpha$-captures.  
   The oxygen production increases with the mixed amount of \nuc{13}C and reaches as large as a half of the upper bound, $Y_{16} / Y_{12} = \sigma_{n \gamma} (\nucm{12}{C}) / \sigma_{n \gamma} (\nucm{16}{O}) \simeq 0.405$ for $\dmixm \gtrsim 0.01$.  
   The amount of oxygen produced is limited by the leak of neutrons from the recycling reactions as they are captured by \nuc{16}O.
   It is followed by the capture of another neutron, while a fraction, $\xi (1-f) / (1 + \xi)$ of neutrons are returned by the $\nucm{17}{O} (\alpha, n) \nucm{20}{Ne} $ reaction.
   Thus, the condition that all the neutrons leave the recycling reactions may be written in the form;
\begin{equation}
\Delta Y_{\rm 13, mix} = \int \sigma_{n \gamma} (\nucm{16}{O}) Y_{16} \left\{ 2 - {\frac{\xi }{(1+\xi)} (1 - f)} \right\} d \tau. \nonumber \\
\end{equation}
   This integral can be carried out with use of eq.~(\ref{eq:o16-sol}), and in particular, for $\tau \ll 1 / \sigma_{n \gamma} (\nucm{16}{O})$, the condition reduces to;
\begin{equation}
	Y_{16}^2  =  Y_{16,0}^2 + {2(1 + \xi) \over 2 + \xi (1 + f) } \left( Y_{12,0} {\sigma_{n \gamma} (\nucm{12}{C}) \over \sigma_{n \gamma} (\nucm{16}{O})} - Y_{16,0} \right) \Delta Y_{\rm 13, mix} . 
\label{eq:o-proa14}
\end{equation}   
     As the amount of mixed \nuc{13}C increases, then, the second term in the right-hand side member overweighs the first term, and the abundance ratio of O to C increases in proportion to a square root of the ratio, \dmix, between the mixed amount, $\Delta Y_{\rm 13, mix}$, of \nuc{13}C to the abundance $Y_{12}$ of \nuc{12}C.
   In this figure, we also plot the average number, $N_{\rm cyc}$, of recycling reactions $\nucm{12}{C} (n, \gamma) \nucm{13}{C} (\alpha, n) \nucm{16}{O}$ that each neutron experiences, estimated by $ N_{\rm cyc} = \delta Y_{16} / \Delta Y_{\rm 13, mix}$. 
   It increases with $\Delta Y_{\rm 13, mix}$ from $N_{\rm cyc} \simeq Y_{\rm 12, 0} \sigma_{n \gamma} (\nucm{12}{C}) / Y_{\rm 16, 0} \sigma_{n \gamma} (\nucm{16}{O})$ to reach a maximum, which is larger by a factor, $(1 + \xi) / (1 + \xi f)$, for larger $\xi $ owing to the contribution of $(\alpha, n)$ reactions, and then, turns to decrease inversely proportional to the square root of \dmix because of the increase of \nuc{16}O.   

   The abundance of \nuc{20}Ne depends on the ratio, $\xi$, of neutron to $\alpha$-capture rates of \nuc{17}O while \nuc{22}Ne is in proportion to mixed amount of \nuc{13}C and $Y_{22} \simeq 0.5 \Delta Y_{\rm 13, mix}$. 
  As a consequence, \nuc{20}Ne can overweigh \nuc{22}Ne for large $\xi $, i.e., for a strong flash of higher temperature, while the neutron capture grows important to convert \nuc{20}Ne into \nuc{22}Ne for a larger exposure of $\tau > 8.4$, and hence, for large mixing amount of $\dmixm > 0.04$.  

\section{Surface Sodium Enhancement during Third Dredge-up}

When the third dredge-up occurs, the carbon abundance in the envelope may greatly increase. 
   A part of the dredged-up carbon is burnt in the hydrogen burning shell to be \nuc{14}N, and then, capture $\alpha$ particles in the helium flash convective zone to be \nuc{22}Ne, which is brought back to the envelope by the succeeding third dredge-up. 
   Furthermore, \nuc{22}Ne, thus enriched, is burnt in the hydrogen burning shell and some is converted to \nuc{23}Na, and again, brought back to the envelope by the third dredge-up. 

If we denote the carbon abundance in the envelope after the $i$-th dredge-up by $Y^{i}_{12}$, the surface abundance of \nuc{22}Ne is increased by the $i$-th dredge-up as much as $\delta Y^{i}_{22} = (Y^{i-1}_{12} / Y_{\rm 12, He, conv}) \delta Y^{i}_{12}$ while the surface carbon abundances is increased by $\delta Y^{i}_{12}$ (where $Y_{\rm 12, He, conv}$ is the carbon abundance in the helium flash convection).  
   The increment of sodium abundance in the envelope is given by $\delta Y^{i}_{23} = g (Y^{i-1}_{22} / Y_{\rm 12, He, conv}) \delta Y^{i}_{12}$, where $g$ is the fraction of \nuc{22}Ne, converted to \nuc{23}Na by the NeNa chain reactions.  
   In the limit of small variations, we may approximate the increases in the abundances by the differential equations, which yields the analytic solutions;
\begin{eqnarray}
Y_{22} & = & Y_{22, 0} + (Y_{12}^2 - Y_{12, 0}^2) / 2 Y_{\rm 12, He, conv} \\
Y_{23} & = & Y_{23, 0} + g \left[ \left( {Y_{22, 0} \over Y_{\rm 12, He, conv}} - {Y_{12,0}^2 \over 2 Y_{\rm 12, He, conv}^2} \right) (Y_{12} - Y_{12, 0}) + {Y_{12}^3 - Y_{12, 0}^3 \over 6 Y_{\rm 12, He, conv}^2} \right].  
\end{eqnarray}
   Here the subscript 0 denotes the abundance at the start of third dredge-up. 

Since the contribution of pristine metals is negligible in the EMP stars, the sodium enrichment of AGB stars may be related to the carbon abundance as;
\begin{equation}
[\abra{Na}{C}] = -2.2 + \log g + 2 [\abra{C}{H}] - 2 \log (Y_{\rm 12, He, conv} / 0.017). 
\label{eq:pro-Natdu} 
\end{equation}
The value of $g$ decreases rapidly with the temperatures in the range relevant to the hydrogen shell-burning; 
  it varies from $g \simeq 1$ for $T \le 5 \times 10^7$ K, to $g \simeq 0.1$ at $T \simeq 6 \times 10^7$ K, and to $g \simeq 0.01$ at $T \ge 7 \times 10^7$ K \citep{Angulo1999}, although the reaction rates of NeNa cycles are subject to fairly large uncertainties \citep[some by more than a order of magnitude][]{ElEid2004}. 

  In the massive AGB stars, carbon is burned in the envelope convection (HBB) to be converted into nitrogen, which enlarges the $[\abra{Na}{C}]$ ratio; 
   this is argued to be the case for stars with EMP stars without the carbon enhancement such as ``mixed stars'' \citep{Spite2005}, where $[\abra{Na}{C}]$ varies from $-0.5$ to 1.0. 
   Furthermore, for still more massive AGB stars, it is possible to burn neon by HBB, which may raise the sodium enrichment up to;  
\begin{equation}
[\abra{Na}{C(+N)}] \simeq -0.07 + \log g +  [\abra{C(+N)}{H}] -  \log (Y_{\rm 12, He, conv} / 0.017). 
\end{equation}

If the temperature during the shell flash reaches above $\sim 3 \times 10^8$ K, on the other hand, most of \nuc{22}Ne is burnt into magnesium in the helium flash convective zone.  
   For such massive stars as enter into TP-AGB phase with the sufficiently large core, we expect the magnesium enrichment as large as 
\begin{equation}
[\abra{Mg}{C}] = -1.4 + [\abra{C}{H}] - \log (Y_{\rm 12, He, conv} / 0.017). 
\label{eq:pro-Mgtdu} 
\end{equation}
In this case, massive isotopes of \nuc{26}Mg and \nuc{25}Mg are enhanced.  

The sodium enrichment is observed at $[\abra{Na}{C}] = - 1.7$ for HE~1327-2326 and at much smaller $[\abra{Na}{C}] = - 3.2$ for HE~0107-5240. 
   In the above equation of eq.~(\ref{eq:pro-Natdu}), $[\abra{C}{H}]$ is the carbon enrichment in the primary stars in our binary scenario and can be larger than currently observed for the low-mass secondary member, polluted via wind accretion.  
   If $[\abra{C}{H}] \simeq 0 - 0.5$, the sodium enrichment of HE0107-5240 can be consistent with that by the third dredge-up alone for $g = 0.1 - 0.01$.
   It is difficult, however, to reproduce the observed sodium enrichment for HE~1327-2326 without hydrogen mixing and He-FDDM. 
   However, the sodium enrichment of HE~0107-5240 can be consistent with the enrichment in eq.~(\ref{eq:pro-Natdu}) with $[\abra{C}{H}] \simeq 0-0.5$, although in this case seek we may not expect the enrichment of aluminum.  
   The rather low enrichments of magnesium observed for HE0107-5240 and for HE~1327-2326 ($[\abra{Mg}{C}] = -3.85$ and $-2.5$, respectively) are indicative of low-mass nature of their erstwhile primary stars. 
   For HE0557-4840, $[\abra{Na}{C}] = - 1.81$ and $[\abra{Mg}{C}] = -1.4$, and yet, the sodium and magnesium are not enriched in comparison with iron ($[\abra{Na}{Fe}] = - 0.16$ and $[\abra{Mg}{Fe}] = 0.25$) as well as aluminum at $[\abra{Al}{Fe}] = - 0.61$ \citep{Norris2007}. 
   It is also noted that HE0557-4840 is distinguishable from HE0107-5240 and HE1327-2326 in the degree of carbon enrichment, which is indicative of the difference in their erstwhile primary AGB stars and in the efficiency of third dredge-up.

\nocite{Aikawa2004}
\bibliography{reference}

\begin{longtable}{cccccccrcr}
\caption{Model Parameters}
\label{tb:para}
\hline
\hline
Model& {$M_{c}^{a}$}  & {$r_{c}^{a}$} & {$\log P^{*b}$} & $\log T_p^{c}$ & $\log L_p^{d}$ & $Y_{12, 0}^{e}$  & $Y_{16, 0}$  & $Y_{12, p}^{f}$ & $Y_{16, p}$ \\
 & {$(M_\odot)$}  & {$(R_\odot)$} & {$({\rm dyn \ cm}^{-2}) $}  & $({\rm K})$ & $(L_\odot)$ &  &  & &  \\
\hline
\endhead
\multicolumn{10}{l}{
\begin{minipage}{10cm}\vspace*{3mm}
a : mass and radius at the bottom of helium burning shell \\
b : proper pressure of the helium shell-flash \\
c : maximum temperature reached during helium shell flashes \\
d : maximum helium burning rate at the peak of shell flashes \\
e : mole abundances of \nuc{12}C and \nuc{16}O at the onset of shell flashes \\
f : mole abundances of \nuc{12}C and \nuc{16}O at the peak of shell flashes
\end{minipage}
}
\endfoot
1 &  0.5713	& 2.046e-2	& 20.001 & 8.295 & 4.33 & 8.758e-3 & 1.141e-5  & 1.205e-2 & 3.596e-5\\
2 &	0.6811	& 1.631e-2	& 19.821 & 8.348 & 4.45 & 1.761e-2 & 4.449e-4	& 2.137e-2 & 4.973e-4 \\
3 &	0.6811	& 1.631e-2	& 20.063 & 8.372 & 5.68 & 1.761e-2 & 4.449e-4  & 2.124e-2 & 4.713e-4 \\
4 &	0.6042	& 1.507e-2	& 20.609 & 8.456 & 7.88 & 1.716e-2 & 2.506e-4	& 2.100e-2 & 2.593e-4 \\
5 &	0.6042	& 1.307e-2	& 20.881 & 8.521 & 9.13 & 1.716e-2 & 2.506e-4	& 2.193e-2 & 2.575e-4 \\
\hline
\end{longtable}

\clearpage

\clearpage

\clearpage

\clearpage

\clearpage

\clearpage

\clearpage

\clearpage

\clearpage

\clearpage

\clearpage

\clearpage

\end{document}